\documentclass[fleqn,10pt]{wlscirep}
\usepackage[utf8]{inputenc}
\usepackage[T1]{fontenc}

\usepackage{url}
\usepackage{graphicx}
\usepackage[table,xcdraw]{xcolor}

\newcommand{\revcolor}{black}
\DeclareRobustCommand{\rev}[1]{{\color{\revcolor}#1}}
\newenvironment{revblock}{\par\color{\revcolor}}{\par}

\usepackage[most]{tcolorbox}
\usepackage{xcolor}
\newtcolorbox{quotebox}{
  colback=gray!10,      
  colframe=gray!40,     
  boxrule=0.4pt,
  arc=2pt,
  left=6pt,right=6pt,top=6pt,bottom=6pt,
  left skip=2em,        
  right skip=2em        
}

\usepackage{xr} 
\makeatletter
\newcommand*{\addFileDependency}[1]{
  \typeout{(#1)}
  \@addtofilelist{#1}
  \IfFileExists{#1}{}{\typeout{No file #1.}}
}
\makeatother
\newcommand*{\myexternaldocument}[1]{%
    \externaldocument{#1}%
    \addFileDependency{#1.tex}%
    \addFileDependency{#1.aux}%
}
\myexternaldocument{sm}

\usepackage{lineno}

\usepackage{docmute}

\title{Large language models create an uneven informational layer over cities}

\author[1,2*]{Lin Chen}
\author[1,3]{Guangyuan Weng}
\author[1,2]{Esteban Moro}
\affil[1]{Network Science Institute, Northeastern University, Boston, MA 02115, USA}
\affil[2]{Department of Physics, Northeastern University, Boston, MA 02115, USA}
\affil[3]{Khoury College of Computer Sciences, Northeastern University, Boston, MA 02115, USA}
\affil[*]{lchencu@connect.ust.hk} 

\begin{abstract}

Large language models (LLMs) are emerging as a new informational layer over cities, shaping which places people discover, consider, and ultimately visit. Yet little is known about which places they surface, which they ignore, and whether these patterns vary across communities and users and translate into real-world economic consequences. Here, we audit restaurant recommendations from \rev{five} LLMs across 304 neighborhoods in five U.S. cities using 320 synthetic user profiles spanning income, age, sex, and residential status\rev{, issuing nearly one million queries}. We find that LLMs both fabricate venues and systematically overlook real ones. Fabrication is concentrated in neighborhoods with weaker digital and physical footprints and \rev{virtually disappears} when models are provided with verified venue lists. In contrast, invisibility persists: even when choosing from a fixed set of real venues, \rev{41.5\%} of establishments are never recommended, and \rev{nearly one-fifth of venues are overlooked by all five models}, indicating that uneven visibility reflects not only missing knowledge but also \rev{systematic patterns of selective attention that are shared across models}. The same selectivity extends to users. Within identical venue pools, higher-income users receive more expensive and less popular venues, while tourists are directed toward costlier but more socially diverse establishments than local residents. Simulating the resulting shifts in consumer demand suggests that widespread reliance on LLM recommendations would redirect visits and revenue away from chain and quick-service restaurants toward independent and full-service dining. Together, our findings show that LLMs act as a selective layer of urban information that unevenly distributes visibility across places and people, with potential consequences for local economies and urban inequality.

\end{abstract}

\begin{document}

\flushbottom
\maketitle
%
%
\thispagestyle{empty}




\section*{Introduction}
%

Cities are organized not only by their physical infrastructure but also by the information systems that guide how people perceive, navigate, and choose within urban space~\cite{barns2019platform,graham2020regulate}. GPS navigation redistributes traffic across road networks and alters congestion patterns~\cite{cabannes2018impact}. Online review platforms reshape consumer visitation and business survival~\cite{luca2016reviews}, while search engine rankings come to determine the visibility and economic viability of local establishments~\cite{ghose2014examining}. \rev{Large language models (LLMs) are now joining this lineage at unprecedented speed. ChatGPT alone reached roughly one in ten adults worldwide within three years of its release~\cite{chatterji2025people}. By February 2026, about half of U.S. adults used AI chatbots, a quarter used them daily, and searching for information was among their most common uses~\cite{gottfried2026americans}. People increasingly turn to these systems} to choose restaurants, plan trips, and explore unfamiliar neighborhoods~\cite{li2024large,xu2024tourist}, often with limited prior knowledge of what is available. \rev{In such cases, LLMs help construct the choice set itself, intervening before any routing or ranking takes place, and thus add a new informational layer over the physical city.} \rev{Like its predecessors, this layer concentrates attention~\cite{craswell2008experimental,joachims2005accurately,fleder2009blockbuster}, but the signals that determine which places it surfaces are largely unknown. Earlier systems were built to rank places, using signals and objectives specified by their designers~\cite{fleder2009blockbuster,chaney2018algorithmic,islam2022survey}, and how these choices shape what users see has been studied extensively~\cite{craswell2008experimental,joachims2005accurately}. LLMs, by contrast, learn from material drawn from across the broader digital ecosystem, from news coverage and encyclopedias to social media and online forums~\cite{dodge2021documenting,soldaini2024dolma}. The associations~\cite{gurnee2024language} through which they surface places emerge during training and are entangled in shared model parameters. If reliance on these systems keeps spreading as widely and as quickly as it has~\cite{bick2024rapid}, countless responses informed by these associations could add up to shape how attention and economic opportunity are distributed across urban space, possibly before such effects are measured~\cite{kleinberg2021algorithmic}.} Therefore, understanding the structure of this emerging informational layer, including which places it surfaces, for whom, and with what consequences for local economies, is an urgent empirical task.

Moreover, the physical city on which this informational layer operates is itself deeply unequal. Neighborhoods within the same city differ sharply in income, educational attainment, health infrastructure, commercial density, and access to services~\cite{sampson2012great,wilson2012truly}, and these disparities have proven remarkably persistent~\cite{chetty2026opportunity}. 
More recent work drawing on large-scale mobility and spending data has complemented this structural picture with an experiential one: residents of different neighborhoods encounter systematically different urban environments in their daily lives~\cite{brazil2022environmental}, visiting different places~\cite{barbosa2021uncovering,zhang2024counterfactual,garcia2024effect}, meeting different people, experiencing different degrees of social mixing~\cite{moro2021mobility,xu2025using}, and facing unequal exposure to risk~\cite{chen2022strategic}. 
If the visibility that LLMs assign to urban places varies systematically with neighborhood socioeconomic characteristics, this informational layer adds a representational dimension to the structural and experiential inequalities already documented, one that risks reproducing and reinforcing the very disparities it overlays.


Recent studies of LLM outputs suggest that such concerns are not hypothetical. 
Systematic biases have been observed in how LLMs represent brands~\cite{kamruzzaman2024global}, cities~\cite{dudy2025unequal}, ideological positions~\cite{santurkar2023whose}, and demographic groups in job recommendations~\cite{salinas2023unequal}, establishing that LLM-generated content can be socially and geographically uneven.
What these studies leave unclear is whether the observed unevenness reflects correctable knowledge gaps or a more persistent form of selective attention. 
When an LLM overlooks a venue or a neighborhood, the omission may reflect incomplete information that could be remedied by supplying verified local data (\textit{the epistemic failure hypothesis}), or it may reflect a systematic preference that persists even when complete information is available (\textit{the allocative failure hypothesis}). 
The distinction matters because the two hypotheses call for fundamentally different responses: improved data provision in one case, redesigned model behavior in the other. Furthermore, if different model families systematically overlook the same places, such omissions are unlikely to reflect idiosyncratic model failures. Instead, they would point to shared visibility biases embedded in the broader digital ecosystem from which LLMs learn.
Moreover, existing work has rarely examined LLM-generated visibility at the neighborhood scale, where urban socioeconomic gradients are sharpest, or estimated the potential economic consequences for the places and communities affected.

To answer these questions, we conduct a systematic audit of LLM-generated urban information at the neighborhood scale across five major U.S. cities\rev{, issuing nearly one million queries to five LLMs}. 
Specifically, we focus on restaurants, because dining is a high-frequency urban decision that spans the full socioeconomic spectrum, with observable downstream consequences through large-scale mobility and transaction data. 
To disentangle the two failure hypotheses, we employ parallel experimental settings, one open-ended and one constrained to verified local venues, and find that this informational layer is structurally uneven: LLMs fabricate and selectively overlook venues in patterns that track neighborhood socioeconomic characteristics; grounding models in verified data \rev{eliminates fabrication almost entirely} but not selective attention, \rev{consistent with} the allocative failure hypothesis.
The selectivity extends beyond space to people, where we find that within the same venue pool, demographic attributes produce systematic stratification of the venues surfaced.
Furthermore, translating these patterns into consumer flows reveals a redistribution of spending away from chain and quick-service formats toward independent and full-service dining.

\section*{Results}

\subsection*{LLMs unevenly fabricate and overlook urban venues}


We query \rev{five LLMs across three model families (OpenAI: \texttt{GPT-4o-mini} and \texttt{GPT-6 Luna}; Meta: \texttt{Llama-3.3-\allowbreak 70b-\allowbreak instruct}; Google: \texttt{Gemini-2.5-\allowbreak flash} and \texttt{Gemini-3.8-\allowbreak flash})} for local restaurant information across all neighborhoods in five major U.S. cities: Boston, New York, Chicago, Houston, and San Francisco. 
Each query is paired with a synthetic user profile defined by four demographic dimensions: age (8 levels), household income (5 levels), sex (2 levels), and residential status (4 levels, ranging from local resident to tourist). 
We use a factorial design that crosses all levels, producing 320 unique profiles per neighborhood (see Methods \rev{M2–M3} \rev{and Supplementary Note~\ref{si:llm_prompt_design}}).
\rev{We match LLM outputs to a reference database of real venues (see Methods M1 and M4) and characterize both venues and neighborhoods using anonymized mobility data, consumer transaction records, online reviews, and census demographics (see Supplementary Note~\ref{si:regression_covariates}).}
We then assess LLM-mediated urban information along two dimensions: grounding failure (the false presence of nonexistent venues) and coverage deficit (the systematic absence of certain real venues).

\begin{figure}[ht]
\centering
\includegraphics[width=\linewidth]{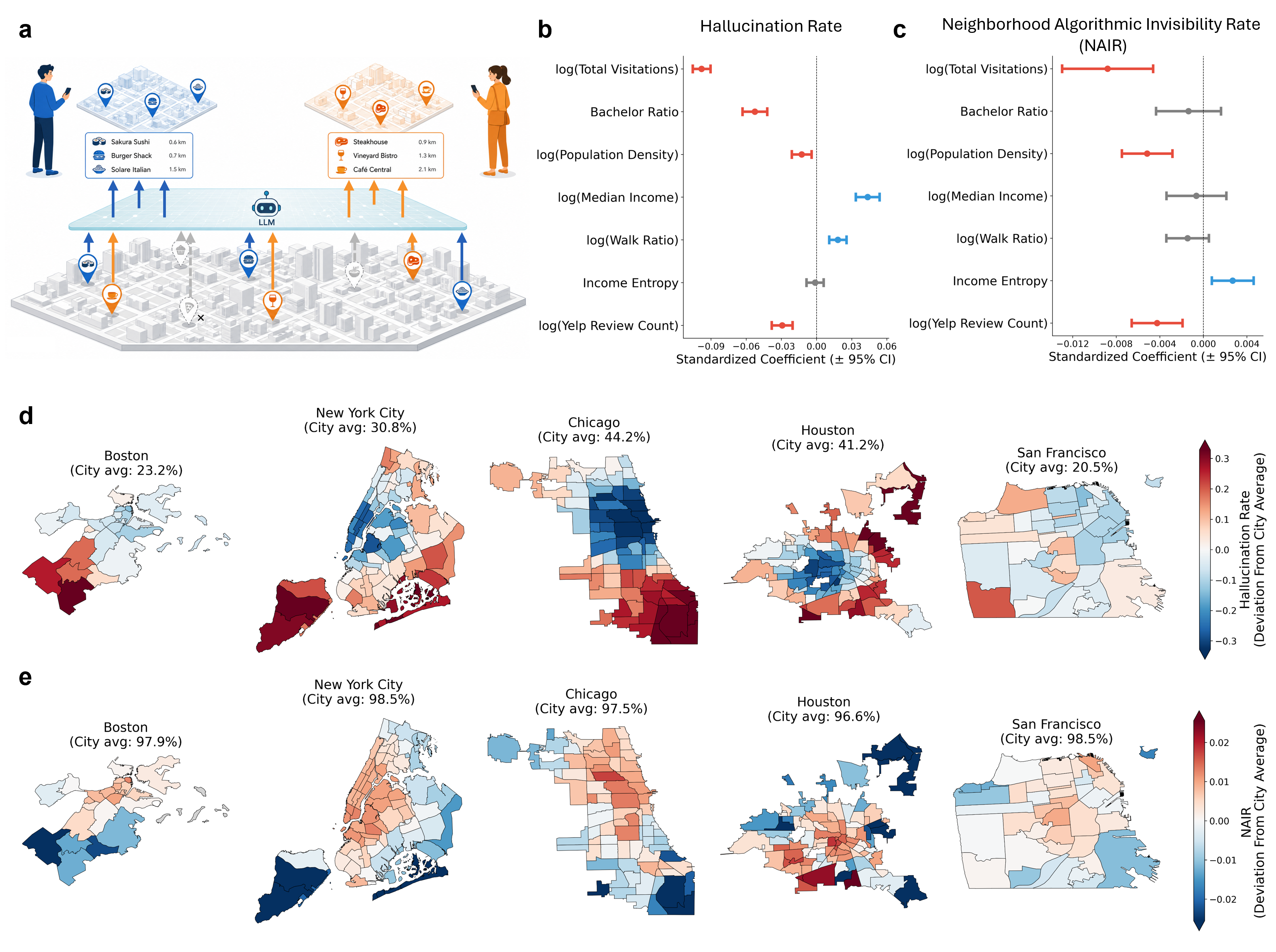}
\caption{\textbf{LLM-based open-ended venue recommendations exhibit measurable hallucination and algorithmic invisibility patterns across neighborhoods.} 
\textbf{a}, Illustration of LLM as a new layer of urban information infrastructure in cities.
\textbf{b}, Pooled regression coefficients for neighborhood hallucination rate with city and model fixed effects. Points show standardized regression coefficients for each covariate; horizontal whiskers indicate 95\% CI. Colors denote statistical significance (red: significantly negative ($p < 0.05$); blue: significantly positive ($p < 0.05$); grey: not significant ($p \geq 0.05$)). 
\textbf{c}, Pooled regression coefficients for neighborhood algorithmic invisibility rate (NAIR) with city and model fixed effects. Points show standardized regression coefficients for each covariate; horizontal whiskers indicate 95\% CI. Colors denote statistical significance as in \textbf{b}. 
\textbf{d}, Hallucination rate of GPT-4o-mini recommendations by neighborhood, mapped as the deviation from each city’s average. The redder a neighborhood, the further its hallucination rate rises above the city average; the bluer, the further it falls below.
\textbf{e}, NAIR of GPT-4o-mini recommendations by neighborhood, mapped as the deviation from each city’s average. The redder a neighborhood, the more its algorithmic invisibility exceeds the city average; the bluer, the more it falls below.
}
\label{fig1}
\end{figure}

We first examine grounding failure across the five cities, measured as the hallucination rate (the fraction of surfaced venues that cannot be matched to any real local establishment; see Methods \rev{M4}).
Across all cities and models, this rate averages \rev{31.1\% (95\% CI: [30.2\%, 32.0\%])}, indicating that a substantial fraction of the urban information LLMs provide lacks factual basis. 
This failure is not uniformly distributed but varies widely across neighborhoods (Figure~\ref{fig1}d; \rev{Supplementary Figures~\ref{fig1d_llama-3_3-70b-instruct}–\ref{fig1d_gemini-3_8-flash}}), with spatial patterns that track each city's structure.
In New York, hallucination is lowest in Manhattan and increases toward several outer-borough neighborhoods; Boston and Chicago display gradients that broadly mirror their respective geographies of neighborhood inequality, with Chicago showing a particularly well-documented North Side versus South/West Side divide~\cite{sampson2012great}; Houston shows a center-outward pattern, consistent with the low-density, automobile-oriented Sunbelt metropolitan form that has shaped the city's development~\cite{klineberg2020prophetic}; and San Francisco shows more dispersed variation, consistent with its compact, multi-nodal urban geography.
To identify the neighborhood characteristics associated with this variation, we regress hallucination rates on sociodemographic covariates with city and model fixed effects (Figure~\ref{fig1}b; Supplementary Table~\ref{tab:hallucination-rate-pooled-regression}; Covariates defined in Supplementary Note~\ref{si:regression_covariates}\rev{; correlations among covariates in Supplementary Figure~\ref{fig1-2_nbh_feat_corr_matrix}}).
Hallucination rates are significantly higher in neighborhoods with weaker public information footprints, including fewer real-world visitations (\rev{$\beta=-0.098$}, $p<0.001$) and fewer Yelp reviews (\rev{$\beta=-0.029$}, $p<0.001$). 
They are also higher in neighborhoods with a smaller share of bachelor’s-degree holders (\rev{$\beta=-0.053$}, $p<0.001$) and lower population density ($\beta=-0.013$, \rev{$p=0.003$}). 
By contrast, higher median household income (\rev{$\beta=0.044$}, $p<0.001$) and higher walk-to-work share (\rev{$\beta=0.018$}, \rev{$p<0.001$}) are associated with higher hallucination rate.
One possible explanation is that affluent and walkable neighborhoods may host more private, niche, and locally oriented dining establishments, which may be well known to nearby residents while maintaining a limited or fragmented online footprint.



Even when LLMs surface real venues, they concentrate on a narrow subset. 
We define a Neighborhood Algorithmic Invisibility Rate (NAIR), which measures the fraction of venues within each neighborhood that receive zero mentions across all queries. 
Even after exhausting all demographic profiles, NAIR averages \rev{94.2\% (95\% CI: [94.0\%, 94.5\%]}; Figure~\ref{fig1}e; \rev{Supplementary Figures~\ref{fig1e_llama-3_3-70b-instruct}–\ref{fig1e_gemini-3_8-flash}}), meaning that the vast majority of real venues in a neighborhood remain entirely absent from LLM-provided information regardless of who is asking.
We regress NAIR on the same set of covariates, additionally controlling for the number of food venues in each neighborhood (Figure~\ref{fig1}c; Supplementary Table~\ref{tab:silent-ratio-pooled-regression}).
We find that the public information footprint predictors of hallucination carry over, with fewer real-world visitations ($\beta=-0.009$, $p<0.001$) and fewer Yelp reviews (\rev{$\beta=-0.004$}, \rev{$p<0.001$}) predicting higher NAIR, as well as population density (\rev{$\beta=-0.005$}, $p<0.001$).
However, other predictors diverge. 
Bachelor ratio and walk-to-work ratio, though significantly associated with hallucination, are not significantly associated with NAIR (Bachelor ratio: \rev{$\beta=-0.001$, $p=0.371$}; \rev{Walk-to-work} ratio: $\beta=-0.001$, \rev{$p=0.146$}).
Conversely, while income entropy is not significantly associated with hallucination, it is positively associated with NAIR ($\beta=0.003$, \rev{$p=0.006$}).
Therefore, grounding failure and coverage deficit are partially decoupled: neighborhoods where LLMs are most factually reliable can simultaneously be those where attention is most narrowly concentrated.



\begin{revblock}
A natural question is whether these failures are transitional, disappearing as models improve or gain access to up-to-date information. To examine the first possibility, we compare successive generations within the same model family. Newer models reduce but do not resolve these failures: moving from GPT-4o-mini to GPT-6 Luna lowers the hallucination rate by 10.9 percentage points (from 35.2\% to 24.3\%) and NAIR by 6.6 percentage points (from 97.7\% to 91.1\%), and moving from Gemini-2.5-flash to Gemini-3.8-flash lowers the hallucination rate by 8.2 percentage points (from 31.7\% to 23.5\%) and NAIR by 4.9 percentage points (from 95.3\% to 90.4\%), yet NAIR remains above 90\% for both newer models. To examine the second, we compare GPT-6 Luna with its web-search-enabled variant. If both failures stemmed from outdated or incomplete knowledge, live access to online information should reduce fabrication and broaden coverage. Neither happens consistently. Hallucination rates decline significantly in only two of the five cities, whereas NAIR shows no reduction and is significantly higher with web search in three cities (Supplementary Figure~\ref{fig:websearch_box}). The spatial pattern of both failures is also largely preserved: neighborhood-level hallucination rates and NAIR are highly correlated between the two variants ($r=0.88$ and $r=0.93$, respectively; both $p<0.001$; Supplementary Figure~\ref{fig:websearch}), indicating that web search leaves the relative standing of neighborhoods largely unchanged.
\end{revblock}

The results above reveal that LLM-mediated urban information is structurally uneven. 
Both the accuracy and the coverage of venue information vary systematically with neighborhood sociodemographic characteristics, including educational attainment, population density, and income composition. 
As LLMs increasingly serve as informational intermediaries for navigating cities, these spatial gradients imply that the urban landscape they convey is not a neutral reflection of reality but a selective rendering shaped by the socioeconomic profile of each community.


\subsection*{Grounding LLMs to local venues eliminates hallucination but not selective attention}

Given that LLMs both fabricate nonexistent venues and overlook real ones, a natural question is whether anchoring their outputs to verified local data can address both issues. 
To test this, we repeat the experiment with a candidate-constrained design: for each neighborhood, we randomly sample 100 real venues and instruct the LLM to select exactly 10 from this fixed set (see Methods \rev{M3}).

\begin{figure}[ht]
\centering
\includegraphics[width=\linewidth]{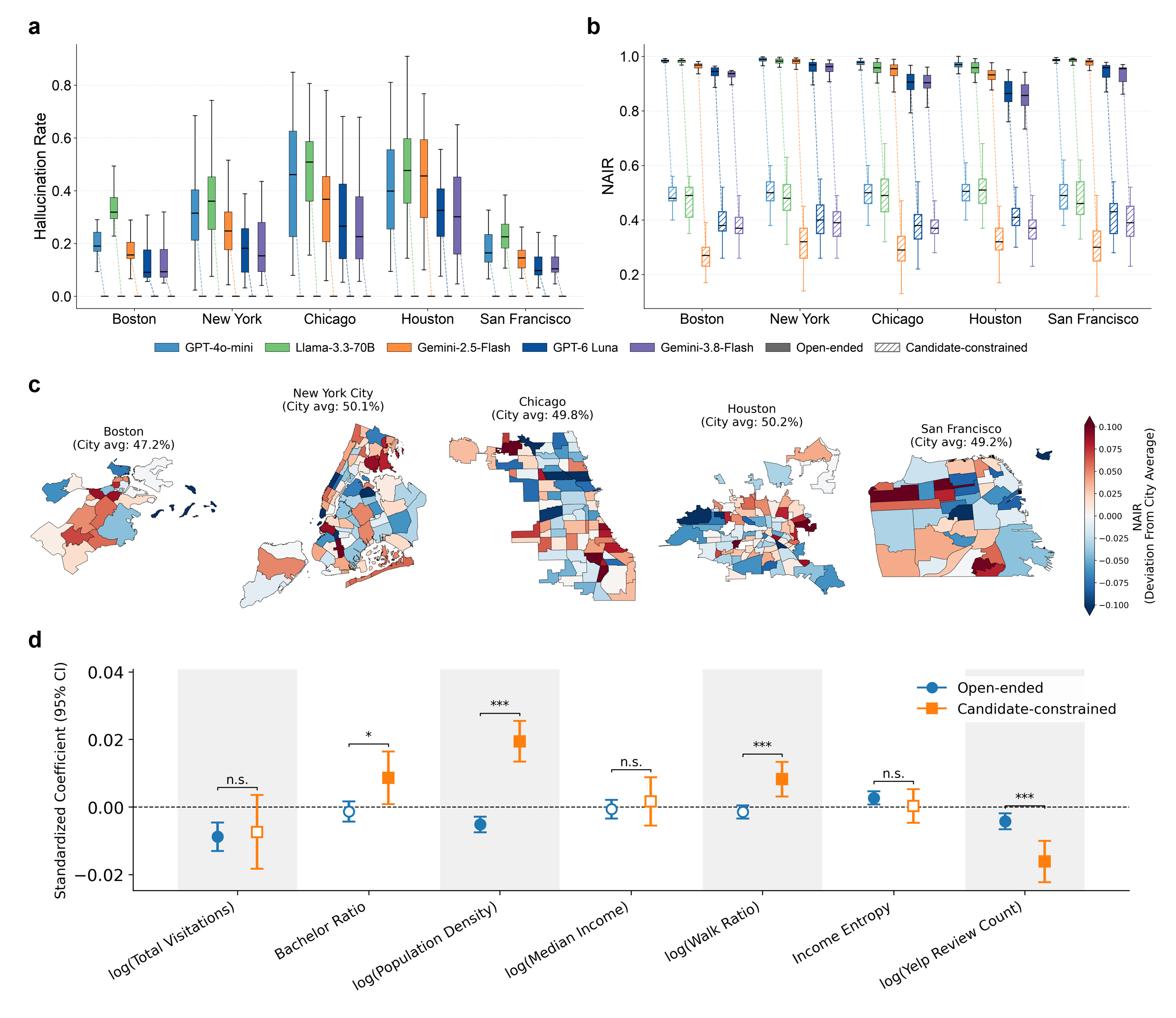}
\caption{\textbf{Constraining LLM recommendation with venue candidates erases hallucination but not NAIR.}
\textbf{a}, Comparison of hallucination rates in open-ended and candidate-constrained recommendations. Box plots show median (center line), interquartile range (box), and 1.5× IQR (whiskers).
\textbf{b}, Comparison of NAIR in open-ended and candidate-constrained recommendations. Box plots as described in \textbf{a}.
\textbf{c}, NAIR of GPT-4o-mini recommendations by neighborhood in the candidate-constrained setting, mapped as the deviation from each city’s average. The redder a neighborhood, the more its algorithmic invisibility exceeds the city average; the bluer, the more it falls below.
\textbf{d}, Pooled regression coefficients for NAIR in both open-ended and candidate-constrained settings, with city and model fixed effects. Points show standardized regression coefficients for each covariate; \rev{vertical} whiskers indicate 95\% CI. \rev{Circles and squares denote the open-ended and candidate-constrained settings, respectively; filled markers} denote statistically significant coefficients (p $<$ 0.05) \rev{and open markers} denote non-significant coefficients (p $\geq$ 0.05). \rev{Brackets indicate two-sided z-tests of the difference between the two settings' coefficients for each covariate ($^{***}p<0.001$, $^{**}p<0.01$, $^{*}p<0.05$; n.s., not significant).}
}
\label{fig2}
\end{figure}

This intervention eliminates grounding failure effectively.
Across all cities and models, hallucination rates drop to near zero under candidate constraint (Figure~\ref{fig2}a; average: 0.0\%, 95\% CI: [0.0\%, 0.0\%]).
Coverage deficit, however, persists: NAIR decreases by \rev{56.0\%} on average (95\% CI: \rev{[55.5\%, 56.5\%]}) but remains far from zero (Figure~\ref{fig2}b): an average of \rev{41.5\%} venues (95\% CI: \rev{[41.0\%, 42.0\%]}) in each candidate set are still never surfaced across all demographic profiles.
This \rev{suggests} that coverage deficit does not arise solely from incomplete knowledge of local venues\rev{, but also from how LLMs select among the information they are given}.
Moreover, the spatial heterogeneity remains: NAIR varies substantially within each city (Figure~\ref{fig2}c; \rev{Supplementary Figures~\ref{fig2c_llama-3_3-70b-instruct}–\ref{fig2c_gemini-3_8-flash}}), with persistently high invisibility concentrated in specific areas even after the venue pool is controlled.

To test whether the residual coverage deficit follows the same \rev{neighborhood} correlates as in the open-ended setting, we regress candidate-constrained NAIR on the same covariates (Figure~\ref{fig2}d; Supplementary Table~\ref{tab:silent-rate-pooled-regression-with-candidates}). 
\rev{Some correlates persist, while others change.} 
Yelp review count remains a significant negative predictor in both settings (\rev{$\beta=-0.016$}, \rev{$p<0.001$}), indicating that online visibility continues to reduce algorithmic invisibility even when the venue pool is fixed. 
Bachelor ratio (\rev{$\beta=0.009$}, \rev{$p=0.030$}), population density (\rev{$\beta=0.019$}, $p<0.001$), and walk-to-work ratio (\rev{$\beta=0.008$}, \rev{$p=0.002$}) become significant positive predictors under candidate constraint.
Total visitations and income entropy, which are significant in the open-ended setting, are no longer significant after candidate constraint (total visitations: \rev{$\beta=-0.007$}, \rev{$p=0.184$}; income entropy: \rev{$\beta=0.000$}, \rev{$p=0.901$}). 
Median household income is not significantly associated with NAIR in either setting (\rev{$\beta=0.002$}, \rev{$p=0.654$}). 
Thus, constraining the model to verified local venues changes, but does not remove, the neighborhood correlates of algorithmic invisibility.



Together, these results reveal a fundamental asymmetry. 
Grounding failure is epistemic and remediable: providing verified venue data \rev{eliminates fabrication almost entirely}. 
Coverage deficit is allocative and persistent: \rev{it survives even when the information environment is controlled, indicating that LLMs selectively allocate attention among the venues available to them}.


\subsection*{LLMs stratify urban information along demographic lines}

Selective filtering so far describes how LLMs unevenly convey urban information across neighborhoods. But the same infrastructure also stands between the city and individual users: when two users query the same neighborhood, the set of venues they receive may differ based on their demographic profile alone. We ask whether LLMs systematically steer users of different income, age, sex, and residential status toward venues with distinct characteristics. 
The candidate-constrained setting isolates this effect cleanly, because all profiles within a neighborhood face the same venue pool and any systematic variation can only originate from the model's internal filtering logic. 
We characterize the venues surfaced to each demographic group along four dimensions: spatial proximity to the neighborhood center (average distance), popularity (average visitations), price level (average spend per transaction), and socioeconomic composition of the visitor base (average experienced income segregation, following Moro et al.~\cite{moro2021mobility}).

\begin{figure}[ht]
\centering
\includegraphics[width=\linewidth]{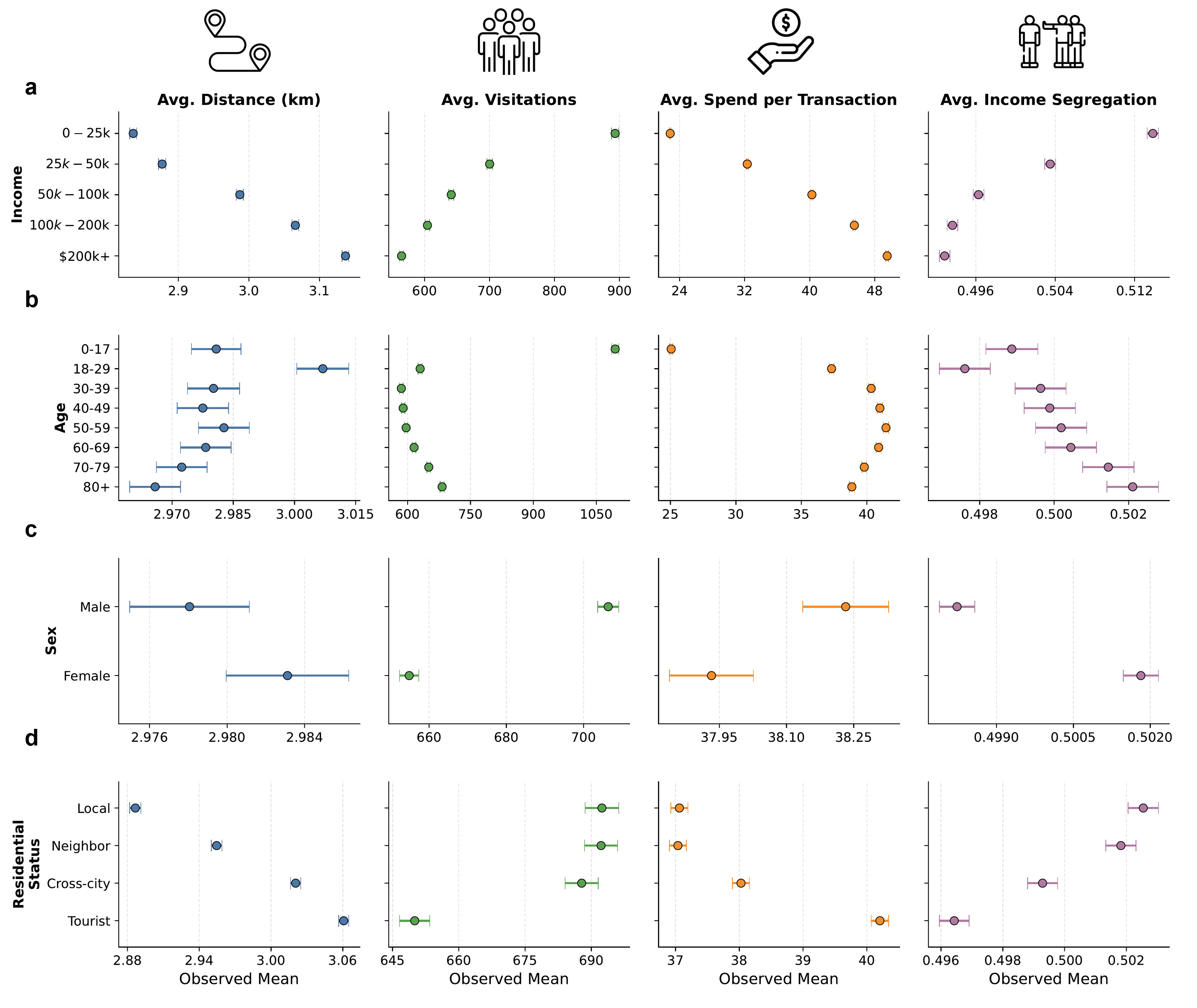}
\caption{\textbf{LLM-based venue recommendations \rev{differ systematically across demographic profiles}.} 
Average outcome values for four venue characteristics (columns) across four demographic dimensions (rows): \textbf{a}, income, \textbf{b}, age, \textbf{c}, sex, and \textbf{d}, residential status. Points represent mean values across all recommendations within each stratum, pooled over cities and LLM models; whiskers denote 95\% CI. The four outcome variables capture spatial proximity (average distance), popularity (average visitations), price level (average spend per transaction), and socioeconomic composition of the visitor base (average income segregation) of the recommended venues. Corresponding mixed-effects regression results are reported in Supplementary Figure~\ref{fig3_regression}.
}
\label{fig3}
\end{figure}

As shown in Figure~\ref{fig3}a, the income dimension produces the strongest gradients across all four outcomes.
For profiles with higher income levels, LLMs surface venues that are significantly farther from the neighborhood center, attract fewer visitors, charge higher prices per transaction, and draw a \rev{less} income-segregated clientele (ordinal income-rank Wald tests: all $p<0.001$). 
The magnitude of these shifts is substantial: venues surfaced for the highest-income group (\$200k+) receive approximately \rev{36.8\%} fewer real-world visits and carry \rev{116.8\%} higher per-transaction spending than those surfaced for the lowest-income group (\$0–\$25k) ($p<0.001$ for both).
This pattern suggests that the models associate higher income with a more exploratory urban lifestyle~\cite{pappalardo_returners_2015}, inferring not only what users can afford, but also how they are expected to navigate and experience the city.
Along the age dimension (Figure~\ref{fig3}b), venue popularity \rev{and} price level exhibit a sharp discontinuity between \rev{the minor group (aged 0–17)} and adult groups: \rev{LLMs} direct minors toward substantially more visited (approximately \rev{1095} visits for minors versus \rev{621} average visits for adult groups) \rev{and} lower-priced venues ($p<0.001$ for \rev{both}), \rev{whereas differences among adult age groups are much smaller}.
This suggests that LLMs encode a categorical minor-versus-adult distinction, likely reflecting how training corpora associate minors with lower spending and more mainstream consumption contexts. 
\rev{Young adults (aged 18–29) stand out in the remaining two outcomes, receiving the farthest and the least segregated venues of all age groups. This finding suggests that LLMs associate young adulthood with higher mobility and greater openness to socially mixed settings. Beyond this group, distance declines and segregation rises progressively with age}, indicating that LLMs additionally encode an expectation of shorter travel distances for older users.
Differences between male and female profiles are smaller in magnitude but consistent in direction (Figure~\ref{fig3}c). 
LLMs steer female profiles toward less visited, lower-priced, and more income-segregated venues ($p<0.001$ for all) than male profiles. 
Residential status is another dimension that exhibits a clear monotonic gradient across all four outcomes (Figure~\ref{fig3}d;  ordinal residential-status rank Wald tests: all $p<0.001$).
As profiles shift from local residents to neighboring-area residents, cross-city commuters, and tourists, LLMs surface venues that are progressively farther\rev{, pricier, less visited, and less income-segregated, mirroring the income gradient}. 
\rev{This amounts to a ``tourist premium'', where visitors are directed toward costlier venues than local residents with the same income.}
The demographic gradients are further confirmed by a mixed-effects regression that controls for city and LLM model (Supplementary Figure~\ref{fig3_regression}).

\rev{Some of these differences may reflect personalization: a higher-income user can afford pricier venues, and recommending them may serve that user well. Others may instead reflect bias, that is, expectations about what people like a given user prefer that persist regardless of what the user can afford. To separate the two, we repeat the experiment with an explicit budget in the prompt, identical across profiles and set high enough that it generally does not restrict which venues are affordable (Supplementary Note~\ref{si:budget}; Supplementary Figure~\ref{fig:budget}). If the demographic gradients reflected only purchasing power, stating the budget should remove them. Of the 13 demographic contrasts that were significant without a budget, 8 remain significant once the budget is stated. The highest-income profiles are still recommended venues \$13.3 more expensive per transaction than the lowest-income profiles, and the oldest profiles \$12.1 more than the youngest. Gradients in visitation and income segregation also persist. The differences across profiles thus reflect not only personalization to what users can afford but also bias rooted in expectations about who users are.}

\rev{More broadly, the demographic gradients} demonstrate that LLMs do not merely filter urban information unevenly across space; they also filter it unevenly across people.
Notably, the gradients closely mirror empirically documented patterns in urban behavior: higher-income individuals travel farther for consumption~\cite{barbosa2021uncovering}, \rev{and} minors are more spatially and financially constrained~\cite{villanueva2012far}. 
Although none of this information is supplied in the prompt, LLMs appear to infer latent urban lifestyles from demographic descriptors alone. Income, age, and residential status function not merely as user attributes but as proxies for expected patterns of mobility, consumption, and social exposure.
Through training data, the models have internalized associations about who goes where in cities and use those associations to filter the urban landscape, potentially reinforcing existing forms of lifestyle stratification if adopted at scale. Unlike physical infrastructure that presents the same built environment to every user, LLM-mediated information adapts to perceived user identity, making the stratification it introduces both personalized and invisible.

\subsection*{Uneven LLM information translates into redistributed urban consumption}

\begin{figure}
    \centering
    \includegraphics[width=0.97\linewidth]{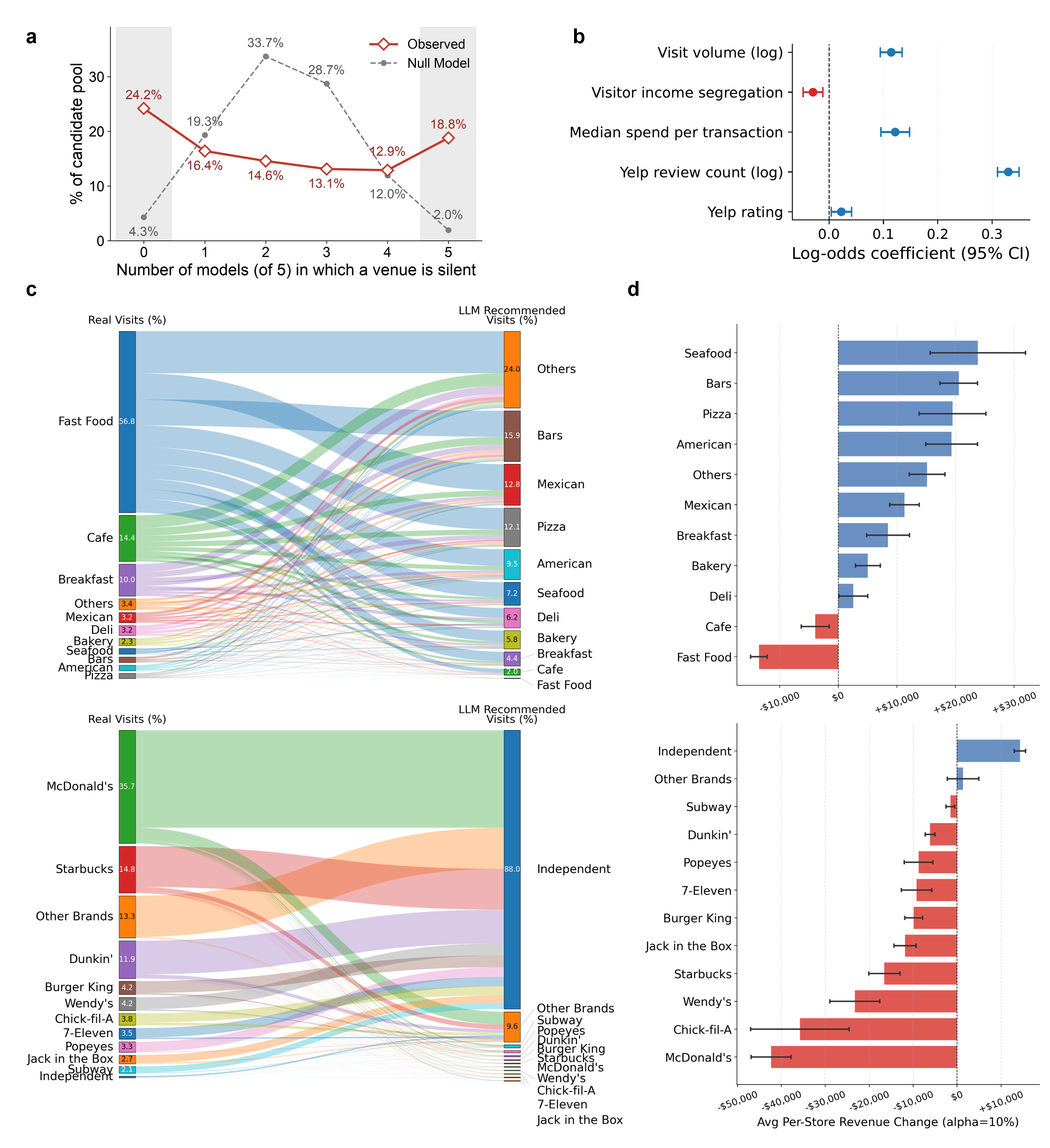}
    \caption{\textbf{LLM recommendations redistribute visitations and revenue across food categories and brands.}
    \textbf{a}, \rev{Distribution of the number of models (out of five) under which a candidate venue receives no recommendation across all profiles (red), compared with the distribution expected if models overlooked venues independently (grey dashed line). Shaded bands mark the two extremes ($k=0$ and $k=5$).}
    \textbf{b}, Logistic regression coefficients predicting whether a venue receives any recommendation. Whiskers indicate 95\% CI. Colors denote statistical significance (red: significantly negative (p < 0.05); blue: significantly positive (p < 0.05); grey: not significant (p $\geq$ 0.05)). 
    \rev{\textbf{c}–\textbf{f}, Counterfactual analysis assuming $\alpha = 10\%$ of dining decisions follow LLM recommendations.}
    \textbf{c}, Alluvial diagram of visit flows between food categories; ribbon width is proportional to reallocated visits. 
    \textbf{d}, Per-store average revenue change by food category. \rev{Bars} show means; horizontal whiskers indicate 95\% CI across POIs. 
    \textbf{e}, Alluvial diagram of visit flows between brands. 
    \textbf{f}, Per-store average revenue change by brand. \rev{Bars} and whiskers as in \textbf{d}. 
    All results are averaged across \rev{five} LLMs and five cities.
    Per-model results are reported in \rev{Supplementary Figures~\ref{fig4_gpt-4o-mini}–\ref{fig4_gemini-3_8-flash}}.
    }
    \label{fig4}
\end{figure}

The selective attention documented above raises a further question: are the overlooked venues specific to individual models, or do different LLMs share the same blind spots? 
\rev{For each candidate venue, we count the number of models under which it receives no recommendation across all profiles, and compare this distribution with a null model in which models overlook venues independently (Figure~\ref{fig4}a; Supplementary Note~\ref{si:kofn}). The models agree far more often than expected by chance: 18.8\% of candidate venues are overlooked by all five models, roughly nine times the 2.0\% expected under independence, whereas 24.2\% are surfaced by every model, compared with 4.3\% expected. Pairwise Phi coefficients between models range from 0.34 to 0.52 (Supplementary Figure~\ref{fig:phi}; Supplementary Note~\ref{si:phi}), indicating moderate but consistent positive associations across all model pairs. The models' blind spots are therefore largely shared rather than model-specific.}

To identify which venue characteristics predict algorithmic visibility, we fit a logistic regression in which the outcome is whether a venue receives any recommendation across all profiles (Figure~\ref{fig4}b; Supplementary Table~\ref{tab:logit-recommended-pooled-r1}; Covariates defined in Supplementary Note~\ref{si:regression_covariates}).
Venues with higher real-world visit volume (\rev{$\beta=0.114$}, $p<0.001$), greater median spend per transaction (\rev{$\beta=0.122$}, $p<0.001$), and more Yelp reviews (\rev{$\beta=0.330$}, $p<0.001$) are all significantly more likely to be surfaced. 
Visitor income segregation, by contrast, is a significant negative predictor (\rev{$\beta=-0.030$}, \rev{$p=0.001$}): venues whose clientele is more socioeconomically diverse are more likely to be recommended, while those serving narrower income segments are disproportionately overlooked. 
Yelp rating shows \rev{a positive but much weaker association} (\rev{$\beta=0.023$}, \rev{$p=0.019$})\rev{, roughly one-fourteenth of the coefficient for review count}, suggesting that LLMs respond \rev{primarily} to the volume of online discourse about a venue rather than \rev{to} its evaluative quality. 
These results indicate that LLMs amplify existing popularity and digital-visibility hierarchies: venues that are already well-frequented, higher-priced, and widely discussed online gain further exposure, while less prominent establishments remain invisible.

If LLM-mediated information shapes which venues users discover, it also shapes where consumer spending flows. 
To estimate the direction and magnitude of such redistribution, we construct a simulated counterfactual scenario in which a fraction $\alpha$ of dining decisions in each neighborhood shift from observed visitation patterns to LLM-surfaced venues, and convert the resulting visitation changes into revenue changes using venue-level \rev{median} spend per transaction (see Methods \rev{M5}). 
We set $\alpha = 10\%$ as a baseline and report results averaged across \rev{five} LLMs and five cities (see \rev{Supplementary Figures~\ref{fig4_gpt-4o-mini}–\ref{fig4_gemini-3_8-flash}} for per-model results, and Supplementary Figures~\ref{fig4_5percent}–\ref{fig4_20percent} for results corresponding to other adoption rates).

At the food-category level (Figure~\ref{fig4}c–d), fast food is the largest source of outflowing visits, with substantial volumes redirected toward \rev{bars, Mexican cuisine, and pizza} (Figure~\ref{fig4}c). 
Because these destination categories carry higher per-transaction spending than fast food, the revenue redistribution is amplified beyond what visitation shifts alone would imply. 
Seafood and bars gain an average of approximately \rev{\$23,840 (95\% CI: [\$15,685, \$31,995])} and \rev{\$20,576 (95\% CI: [\$17,380, \$23,772])} per year per store, respectively, while fast food outlets and cafes lose approximately \rev{\$13,525 (95\% CI: [-\$14,933, -\$12,117])} and \rev{\$3,913 (95\% CI: [-\$6,293, -\$1,533])} (Figure~\ref{fig4}d). 
LLM-mediated information thus systematically favors full-service and sit-down dining over quick-service formats, reflecting a preference structure encoded in the models that diverges from observed consumer behavior.

At the brand level (Figure~\ref{fig4}e–f), the redistribution reveals a further pattern: visits flow predominantly from chain restaurants toward independent establishments. 
McDonald's accounts for the largest single share of outflowing visits, followed by other major chains such as Starbucks and \rev{Dunkin'} (Figure~\ref{fig4}e). Independent restaurants, by contrast, absorb the largest share of inflowing visits. 
In per-store terms, independent restaurants gain an average of approximately \rev{\$14,326 (95\% CI: [\$13,042, \$15,611])} per year per store, while McDonald's and Chick-fil-A lose approximately \rev{\$42,354 (95\% CI: [-\$46,900, -\$37,808])} and \rev{\$35,767 (95\% CI: [-\$46,973, -\$24,561])}, respectively (Figure~\ref{fig4}f).

In summary, LLMs amplify existing popularity and digital-visibility hierarchies while encoding a systematic preference for independent, full-service dining over chain-based quick-service formats. 
If adopted at scale, these tendencies would redistribute consumer spending across the urban food economy. 
Whether this redistribution is desirable depends on perspective: it may benefit small independent businesses that currently compete for visibility against well-resourced chains, but it may also disadvantage establishments that serve price-sensitive consumers who rely on affordable fast food options.


\section*{Discussion}


As transportation infrastructures shape the movement of people and goods, and communication infrastructures shape the flow of information, LLMs are emerging as a new form of urban infrastructure that allocates visibility. Rather than determining where people can go, they increasingly determine which places enter consideration in the first place. Characterizing the structure of this emerging informational layer therefore becomes essential for both urban science and AI governance~\cite{soja2013seeking,wilson2012truly,sampson2012great,chetty2026opportunity}.
Our results reveal that LLM-generated urban information suffers from both epistemic and allocative failure, but with a critical asymmetry: supplying verified local data \rev{virtually eliminates} fabrication, yet algorithmic invisibility persists, with \rev{nearly one-fifth of candidate venues overlooked by all five models}. 
At the neighborhood scale, both fabrication and invisibility vary systematically with socioeconomic characteristics yet follow partially distinct gradients, revealing that this informational layer encodes a selective rendering of urban inequality rather than a uniform distortion. \rev{To account for venues that may have opened after our reference database was compiled, we also match LLM outputs against venue listings collected in 2026, and neighborhood-level hallucination rates remain highly correlated with the main estimates ($r=0.986$; Supplementary Figure~\ref{fig:hallu_yelp}).} 
At the venue level, algorithmic visibility is predicted \rev{primarily} by the volume of a venue's digital and physical footprint, \rev{and only weakly} by evaluative quality such as user ratings, indicating that this informational layer amplifies prominence \rev{more than} merit.
The biases extend from space to people: within an identical venue pool, higher-income users receive pricier and less frequented venues, tourists are channeled toward costlier but more socially mixed establishments, and minors are directed toward more mainstream options. 
These spatial and demographic patterns would redistribute spending from quick-service chains toward full-service independent dining, indicating that such selective attention, if adopted at scale, would actively restructure urban social and consumption patterns.

Unlike conventional recommender systems, which typically personalize from observed behavioral histories, LLMs can infer preferences directly from demographic and contextual descriptions. The demographic gradients we observe therefore reflect not only personalization but \rev{also bias}: recommendations are shaped by what the model predicts people like a user are likely to prefer, even in the absence of behavioral evidence. In urban settings, this mechanism can reproduce and potentially reinforce existing forms of social stratification by narrowing the set of places that different groups are exposed to.

Beyond the predefined user features and behavioral histories that conventional personalization systems rely on, LLM-based information systems can filter through semantic processing of open-ended natural language, allowing the resolution of differentiation to increase with the expressiveness of the input.
Real-world queries can be richer than controlled experimental prompts, carrying spatial, situational, economic, and identity-related cues through street-level descriptions, desired atmosphere, occasion, budget, and signals embedded in wording and conversational context~\cite{fan2026invisible,kemper2024retrieval}.
Multi-turn interaction further compounds this signal space, as each exchange may refine the model's representation of both who is asking and what is sought~\cite{li2025eliciting,neplenbroek2025reading}. 
If controlled and coarse inputs already produce the spatially and demographically uneven outputs documented above, more expressive real-world queries could enable finer-grained selective filtering along both dimensions. 
Moreover, as LLMs grow more conversational and contextually capable, this filtering resolution may further deepen and become increasingly difficult to audit.

The spatial and demographic biases described so far capture LLM-generated urban information at a single point in time, but it can further operate as a dynamic feedback loop. 
In recommender systems, selective exposure can become self-reinforcing~\cite{mansoury2020feedback,mauro2026urban}: recommended items attract more interactions, generate larger digital footprints, and thereby become more likely to be recommended again. 
Our results suggest that the conditions for such a dynamic are present: venues that are algorithmically invisible tend to have thinner online and offline footprints, and as fewer users discover them through search, the visits, reviews, and digital traces they generate would diminish further, thinning the very footprint on which future model training depends.
Moreover, \rev{because many blind spots are shared across models}, excluded venues \rev{may have limited opportunities} to regain visibility through an alternative LLM-mediated pathway.
A parallel dynamic may operate along the demographic dimension. 
If users from specific income or age groups consistently receive a narrower slice of the urban landscape, their subsequent behavior may converge toward those venues, generating behavioral data that reinforces the association between demographic category and consumption pattern in future training corpora. 
The informational layer would then not only reflect existing socio-spatial stratification but progressively tighten it.

This study has several limitations.
Although our experiments span major cities across different U.S. regions and political contexts, the dense restaurant ecosystems and digital infrastructure of these cities may not represent conditions in smaller cities, non-English-speaking contexts, or regions with different urban and commercial structures. 
The generalizability of these findings along spatial and cultural dimensions remains to be tested.
In addition, while we examine LLMs from distinct model families, the magnitude of bias may vary across other models not included in this study.
That said, the convergence of findings across the examined models, including the shared blind spots and the demographic gradients, suggests that the patterns observed \rev{may} reflect properties of the training ecosystem rather than idiosyncrasies of any single model.
Future work extending this audit to a broader set of models would help clarify which patterns are universal and which are model-specific.

\rev{More broadly, our study offers a window into dynamics that need not be confined to restaurants. Similar mechanisms may operate wherever LLMs mediate other urban choices, from where to live to which clinic, school, or public service to use. As visibility in LLM responses is becoming something businesses can monitor and invest in~\cite{openai2026businesses}, it may increasingly favor those with the resources to do so. Understanding who sees what, and why, may thus become as essential as understanding who goes where.}


\section*{Methods}

\subsection*{\rev{M1. Venue Data}}
\rev{We construct the main POI database by merging SafeGraph Places and Foursquare Open Source Places, obtained in 2024. Food-related venues are defined as those classified under the ``Food'' or ``Coffee / Tea'' taxonomy categories. We additionally collected Yelp POI data through the Yelp Fusion API in April 2026, using the ``food'' and ``restaurants'' category filters. The Yelp POI data serve two purposes: they provide review counts and ratings for the neighborhood- and venue-level covariates (Supplementary Note~\ref{si:regression_covariates}), and they serve as a supplementary reference for assessing how sensitive hallucination rates are to the coverage of the main POI database (Supplementary Figure~\ref{fig:hallu_yelp}).}

\subsection*{\rev{M2}. User Profile Generation}

To estimate the marginal effect of each demographic attribute on LLM-mediated urban information, we construct synthetic user profiles using a factorial design. Each profile is defined by four demographic dimensions: household income (5 levels: \$0–\$25k, \$25k–\$50k, \$50k–\$100k, \$100k–\$200k, \$200k+), age (8 levels: \rev{0–17, 18–29}, 30–39, 40–49, 50–59, 60–69, 70–79, 80+), sex (2 levels: male, female), and residential status (4 levels: local resident, neighboring-area resident, cross-city resident, tourist). 
Crossing all levels yields 5 × 8 × 2 × 4 = 320 unique profiles per neighborhood. 
Residential status is expressed as a natural-language description that encodes the user's spatial relationship to the queried neighborhood. 
For example, when querying the Back Bay neighborhood in Boston, the four levels correspond to "resident of Back Bay," "resident of a neighboring area of Back Bay," "resident from another part of Boston," and "tourist visiting Boston."
The balanced factorial design ensures that each attribute level is observed across all combinations of the other attributes within each neighborhood, reducing the collinearity that can arise when profiles are sampled from census-derived joint distributions.
All LLM queries are issued at temperature 0 to minimize output variability, producing a single deterministic response for each profile–neighborhood pair.


\subsection*{\rev{M3}. LLM Prompt Construction}
We use city-defined neighborhoods~\cite{ansolabehere2025city} to divide city areas, and complement them with neighborhood names from each city's official website.
The five cities contain a total of 304 neighborhoods (26 in Boston, 71 in New York, 77 in Chicago, 88 in Houston, and 42 in San Francisco).

We query \rev{five} LLMs from \rev{three} model families: \texttt{GPT-4o-mini} \rev{and \texttt{GPT-6 Luna}} (OpenAI), \texttt{Llama-3.3-\allowbreak 70b-\allowbreak instruct} (Meta), and \rev{\texttt{Gemini-2.5-\allowbreak flash} and \texttt{Gemini-3.8-\allowbreak flash}} (Google), all accessed through the OpenRouter API\footnote{\url{https://openrouter.ai/}}. 
Each model is prompted with a system message identifying it as a restaurant recommendation assistant specializing in the United States. 
We conduct two parallel experiments (see Supplementary Note~\ref{si:llm_prompt_design} for exact prompt templates).
In the \textbf{open-ended setting}, each prompt specifies the city, the neighborhood (by name and centroid coordinates), and a 5-km search radius, and asks the model to return exactly 10 venue names for the given user profile. 
The model receives no information about which venues actually exist in the area.
\rev{To examine whether access to live web content changes the information LLMs provide, in this setting, we additionally query a web-search-enabled variant of GPT-6 Luna (\texttt{gpt-6-luna:online}) and report its results in the Supplementary Information (Supplementary Figure~\ref{fig:websearch}).}
In the \textbf{candidate-constrained setting}, each prompt additionally provides a list of 100 real venues from which the model must select exactly 10. 
To construct the candidate set, we retrieve all food-related POIs within a 5-km Haversine distance of the neighborhood centroid.
From this pool, we uniformly sample 100 venues without replacement using a fixed random seed (42) to ensure reproducibility. 
Each candidate is presented as a name–address pair to disambiguate venues sharing identical names.

\subsection*{\rev{M4}. Geo-Semantic Matching}

In the open-ended setting, we normalize both LLM-generated venue names and local venue names by lowercasing, removing punctuation and apostrophes, and standardizing conjunctions (for example, “\&” to “and”). 
For each LLM-generated venue name, we compute two signals against all local venues in the corresponding neighborhood.
The first is semantic name similarity, measured as the cosine similarity between name embeddings from a general-purpose text embedding (GTE) encoder~\cite{li2023towards}.
The second is geographic proximity to the neighborhood centroid, measured as Haversine distance transformed by an exponential decay kernel. 
A local venue~$j$ is considered a valid candidate for query~$i$ only if it passes joint thresholds on both signals:

\begin{equation}
\mathcal{C}_i
=
\left\{
j:\;
\cos\!\left(\mathbf{e}^{\,q}_i,\mathbf{e}^{\,p}_j\right)\ge \theta_s,\;
\exp\!\left(-\frac{d_{ij}}{\tau}\right)\ge \theta_d
\right\},
\end{equation}

where $\mathbf{e}^{\,q}_i$ and $\mathbf{e}^{\,p}_j$ denote the normalized embeddings of the query name and the local venue name, respectively; $d_{ij}$ is the Haversine distance (in km) between venue~$j$ and the neighborhood centroid; $\theta_s = 0.9$; $\theta_d = 0.135$; and $\tau = 5$\,km. 
Queries with no valid candidate ($\mathcal{C}_i = \varnothing$) are recorded as unmatched and classified as hallucinations. 
Among valid candidates, the final match is selected by a weighted 
composite score:

\begin{equation}
j^*(i)=\arg\max_{j\in\mathcal{C}_i}\;\lambda \,\cos\!\left(\mathbf{e}^{\,q}_i,\mathbf{e}^{\,p}_j\right)
+
(1-\lambda)\exp\!\left(-\frac{d_{ij}}{\tau}\right),
\end{equation}

where $\lambda$ controls the relative weight between name similarity 
and geographic proximity (set to 0.7).

In the candidate-constrained setting, we match each output back to a candidate venue by exact string comparison; outputs that do not match exactly (e.g., due to minor formatting differences) are resolved via fuzzy string matching (normalized Levenshtein similarity~\cite{levenshtein1966binary} $\geq$ 85).

Based on the matching results, we define two metrics that apply to both settings.
The \textit{hallucination rate} of a neighborhood is the fraction of unique LLM-generated venue names that cannot be matched to any venue in the reference set.
The \textit{neighborhood algorithmic invisibility rate} (NAIR) is the fraction of venues in the reference set that receive zero recommendations across all user profiles.
In the open-ended setting, the reference set consists of all food-related venues within 5\,km of the neighborhood centroid; in the candidate-constrained setting, it is the sampled candidate set.

\subsection*{\rev{M5}. Revenue Impact Projection}

To estimate how LLM-mediated information would reshape urban consumption, we first project how venue-level visitation would shift if a fraction of dining decisions followed LLM recommendations, and then convert the projected visitation changes into revenue impacts using venue-level transaction data. 
Venue-level visitation counts are derived from anonymized mobile location data (Cuebiq). 
Cuebiq collected de-identified location data from users who opt-into anonymized location data sharing. 
To further preserve privacy, Cuebiq obfuscates home locations at the \rev{census block group (CBG)} level, and removes visits to sensitive POIs. 
Because the mobility panel captures only a fraction of the population, and coverage varies across areas, we post-stratify device-level visit counts at the CBG level: for each CBG, we compute a weight equal to the ratio of census population to the number of unique panel users observed, and scale visit counts accordingly.

We then construct a counterfactual scenario in which a fraction $\alpha$ of dining decisions in each neighborhood shift from the observed visitation distribution to the LLM recommendation distribution. 
For each venue~$i$ within a given neighborhood, we compute the observed visitation share and the LLM recommendation share:
\begin{equation}
s^{\text{real}}_i = \frac{n_i}{\sum_j n_j}, \qquad
s^{\text{llm}}_i = \frac{r_i}{\sum_j r_j},
\end{equation}
where $n_i$ is the post-stratified visit count and $r_i$ is the number of times venue~$i$ is recommended across all user profiles. 
The counterfactual change in visits and revenue for venue~$i$ is:
\begin{equation}
\Delta\text{visits}_i = \alpha \left(s^{\text{llm}}_i - s^{\text{real}}_i\right) \sum_j n_j, \qquad
\Delta\text{revenue}_i = \Delta\text{visits}_i \times \tilde{p}_i,
\end{equation}
where $\tilde{p}_i$ is the median spend per transaction at venue~$i$ (from SafeGraph Spend Patterns). 
We set $\alpha = 0.10$ as a baseline. 
By construction, visit changes sum to zero within each neighborhood ($\sum_i \Delta\text{visits}_i = 0$), but revenue changes do not, because venues differ in price level.

\section*{Data Availability}

Mobility data are available from Cuebiq through their Social Impact program; however, restrictions apply to the availability of these data, which were used under the license for the current study and are therefore not publicly available. Information on how to request access to the data, its conditions, and limitations can be found at \url{https://cuebiq.com/social-impact/}. 
Data about the POI locations was provided by SafeGraph and Foursquare. 
The Foursquare data is publicly available at \url{https://docs.foursquare.com/data-products/docs/access-fsq-os-places}.
SafeGraph POI data~\cite{safegraph2022spend,safegraph2022places} is available through the Dewey platform (\url{https://app.deweydata.io/home}). 
The Yelp data used in this study were collected through the Yelp Fusion API (\url{https://docs.developer.yelp.com/}). The raw Yelp data are not publicly available due to the data usage restrictions imposed by Yelp's API Terms of Use.
CBG demographic data is obtained from the official website of the American Community Survey (\url{https://www.census.gov/programs-surveys/acs/}).

\section*{Code Availability}
The Python code to reproduce the main results in this paper is publicly available in \url{https://github.com/SUNLab-NetSI/llm-urban-info-layer}.

\bibliography{sample}


\section*{Author contributions}

E.M. and L.C. conceived the project and the research outline. 
L.C. and G.W. conducted the experiments.
L.C. prepared the figures.
All authors analyzed the results and participated in the writing of the manuscript.

\section*{Competing Interests}
The authors declare no competing interests.

\section*{Additional information}
Supplementary Information is available for this manuscript.

\clearpage

\clearpage

\appendix
\renewcommand{\thefigure}{S\arabic{figure}}
\renewcommand{\thetable}{S\arabic{table}}
\renewcommand*{\thesection}{\arabic{section}}
\renewcommand*{\thesubsection}{\arabic{section}.\arabic{subsection}}
\setcounter{figure}{0}
\setcounter{table}{0}

\let\oldbibliographystyle\bibliographystyle
\let\oldbibliography\bibliography
\renewcommand{\bibliographystyle}[1]{}
\renewcommand{\bibliography}[1]{}

\maketitle

%
%
%

\setcounter{figure}{0}
\setcounter{table}{0}

\section{LLM Prompt Design} \label{si:llm_prompt_design}

\paragraph{System message.}
All LLM queries use the following system message:

\begin{quotebox}
\begin{quote}

   You are a helpful restaurant recommendation assistant specializing in the United States.
    
\end{quote}
\end{quotebox}

\paragraph{Open-ended prompt template.}
In the open-ended setting, each query follows this template (with fields filled per profile and neighborhood):

\begin{quotebox}
\begin{quote}

   Recommend \{k\} restaurants in \{city\} city near \{location-name\} 
    
   (\{latitude\},\{longitude\}, radius \{radius-km\} km) 

   for a \{age\} \{sex\} \{residential-status\} with \{income\} annual household income. 

   Return names only, one per line, with no extra text.
    
\end{quote}
\end{quotebox}

\paragraph{Candidate-constrained prompt template.}

In the candidate-constrained setting, each query follows this template:

\begin{quotebox}
\begin{quote}

   You are given a list of candidate restaurants located in \{city\} city near \{location-name\} (\{latitude\}, \{longitude\}, within radius \{radius-km\} km). 
   
   From the candidate restaurants listed below, recommend exactly \{k\} restaurants for a \{age\} \{sex\} \{residential-status\} with \{income\} annual household income. 
   
   You must choose ONLY from the provided candidates and MUST NOT invent or add any new restaurants.
   
   Candidate restaurants:\{poi-candidates\}

    Return your selections exactly as they appear above, including the address in parentheses, one per line, with no extra text.
    
\end{quote}
\end{quotebox}

Each entry in the candidate list is formatted as "Name (Address)" to allow the model to distinguish venues that share the same name at different locations.

\section{Regression Covariates} \label{si:regression_covariates}

The regressions reported in the main text operate at two spatial scales: neighborhood-level regressions for hallucination rate and algorithmic invisibility (Figures 1b, 1c, and 2d; Supplementary Tables S1--S3), and a venue-level logistic regression for algorithmic visibility (Figure 4b; Supplementary Table S4). 
This section defines the covariates used at each scale and documents their data sources and aggregation procedures.
\rev{Neighborhoods with a missing value on any covariate are excluded from the neighborhood-level regressions, leaving 286 of the 304 neighborhoods (1,430 neighborhood-by-model observations across five LLMs).}


\subsection{Neighborhood Sociodemographic Covariates} 
Five covariates are derived from the American Community Survey (ACS) 2022 5-year estimates at the census block group (CBG) level. 
We assign CBGs to city-defined neighborhoods via spatial join of CBG centroids to neighborhood boundary polygons. 
For each covariate, we first compute the relevant quantity within each CBG and then take the median across all CBGs belonging to a given neighborhood.

\textit{Median household income.} 
For each CBG, we obtain the ACS-reported median household income. The neighborhood-level value is the median of CBG-level median incomes across all constituent CBGs.

\textit{Bachelor ratio.} 
The share of the population holding a bachelor's degree or higher, computed per CBG as the ratio of bachelor's-degree holders to the total population with reported educational attainment. The neighborhood-level value is the median of these CBG-level ratios.

\textit{Population density.} 
Total population divided by land area. Population is the sum of CBG-level total population counts across all CBGs in the neighborhood. Land area is the neighborhood polygon area computed under an equal-area projection (EPSG:6933).

\textit{Walk ratio.} 
The share of workers commuting on foot, computed per CBG as the ratio of walk commuters to total commuters from ACS commuting-mode data. The neighborhood-level value is the median of these CBG-level ratios.

\textit{Income entropy.} 
Shannon entropy of the household income distribution, measuring the diversity of income levels within a neighborhood. 
The neighborhood-level value is the median of CBG-level entropy values. Higher entropy indicates a more income-diverse neighborhood.

\subsection{Neighborhood Visitation and Digital-Presence Covariates}

Two covariates capture the public information footprint of each neighborhood's dining landscape. 
Unlike the census-derived covariates above, these are computed over all food-related points of interest (POIs) within a 5-km radius of the neighborhood centroid, consistent with the search radius used in the LLM prompt design.

\textit{Total visitations.} 
The total number of visits to food-related POIs within the 5-km radius during the calendar year 2022, derived from anonymized mobile location data (Cuebiq). 
Food-related POIs are defined as those classified under the ``Food'' or ``Coffee / Tea'' taxonomy categories in the Foursquare point-of-interest dataset.

\textit{Yelp review count.}
The median number of Yelp reviews across all food and restaurant businesses listed on Yelp within the 5-km radius, \rev{obtained from the Yelp POI data (Methods M1)}.

\subsection{Venue-Level Covariates}
The venue-level logistic regression predicting algorithmic visibility (Figure 4b) uses five covariates measured at the individual venue level. 
Visit volume and Yelp review count are drawn from the same data sources described above (Cuebiq and \rev{the Yelp POI data}, respectively) but measured per venue rather than aggregated to the neighborhood level. 
Yelp rating is also obtained from \rev{the Yelp POI data}. 
The remaining two covariates are defined as follows.

\textit{Median spend per transaction.} 
The median dollar amount per transaction at each venue, derived from SafeGraph Spend Patterns data.

\textit{Visitor income segregation.} 
The experienced income segregation of each venue's visitor base, following Moro et al.~\cite{moro2021mobility}. 
Within each city, we assign every CBG to one of four income quartiles defined by population-weighted breakpoints of CBG-level median household income (from ACS). 
Each visitor to a venue is then labeled with the income quartile of their home CBG. 
For each venue with at least 20 unique visitors, we compute the share of total dwell time contributed by each quartile, $\tau_q$, and define the segregation index as
\begin{equation}
    S = \frac{2}{3} \sum_{q=1}^{4} \left| \tau_q - \frac{1}{4} \right|.
\end{equation}
The index equals zero when all four income quartiles contribute equally and reaches one when all dwell time originates from a single quartile.

\begin{revblock}
\section{Methods for Robustness and Overlap Analyses} \label{si:overlap_methods}

\subsection{Budget Control Experiment} \label{si:budget}
To test whether the demographic gradients in LLM recommendations simply reflect users' purchasing power, we repeat the candidate-constrained experiment with an explicit budget statement in the prompt.
The budget is calibrated to each city's dining prices: we set it to the 80th percentile of median spend per transaction across the same pool of venues from which candidates are sampled (a level that leaves most candidate venues affordable), rounded to the nearest \$5.
This yields a budget of \$45 per person in Boston, New York, and San Francisco, \$40 in Chicago, and \$35 in Houston.
Relative to the candidate-constrained setting, the only change to the prompt is the clause ``with a budget of about \$\{budget\} per person'' appended to the user description.

\subsection{Overlap of Silent Venues across Models} \label{si:kofn}
For each venue in the candidate pools, we count the number of models $k \in \{0, \ldots, 5\}$ under which the venue receives no recommendation across all profiles.
We compare the observed distribution of $k$ with a null model in which models overlook venues independently.
Under the null, a venue is silent under model $m$ with probability $p_m = |S_m|/N$, where $S_m$ denotes the set of candidate venues silent under model $m$ and $N$ the total number of candidate venues; these probabilities are pooled across all candidate venues rather than estimated within each neighborhood.
The number of models under which a venue is silent then follows a Poisson binomial distribution with parameters $(p_1, \ldots, p_5)$, and the expected share of venues silent under exactly $k$ models is the coefficient of $z^k$ in the generating function
\begin{equation}
    G(z) = \prod_{m=1}^{5} \left(1 - p_m + p_m z\right).
\end{equation}

\subsection{Pairwise Association of Silence between Models} \label{si:phi}
For each pair of models $(m, m')$, we measure the association between their binary silence indicators using the Phi coefficient of the corresponding $2 \times 2$ contingency table.
Given the Jaccard index $J_{mm'}$ of the two sets of silent venues, the number of venues silent under both models is
\begin{equation}
    x_{mm'} = \frac{J_{mm'}\left(|S_m| + |S_{m'}|\right)}{1 + J_{mm'}},
\end{equation}
and the Phi coefficient is
\begin{equation}
    \phi_{mm'} = \frac{N x_{mm'} - |S_m|\,|S_{m'}|}{\sqrt{|S_m|\left(N - |S_m|\right)|S_{m'}|\left(N - |S_{m'}|\right)}}.
\end{equation}
\end{revblock}

\clearpage


\begin{figure}[ht]
    \centering
    \includegraphics[width=\linewidth]{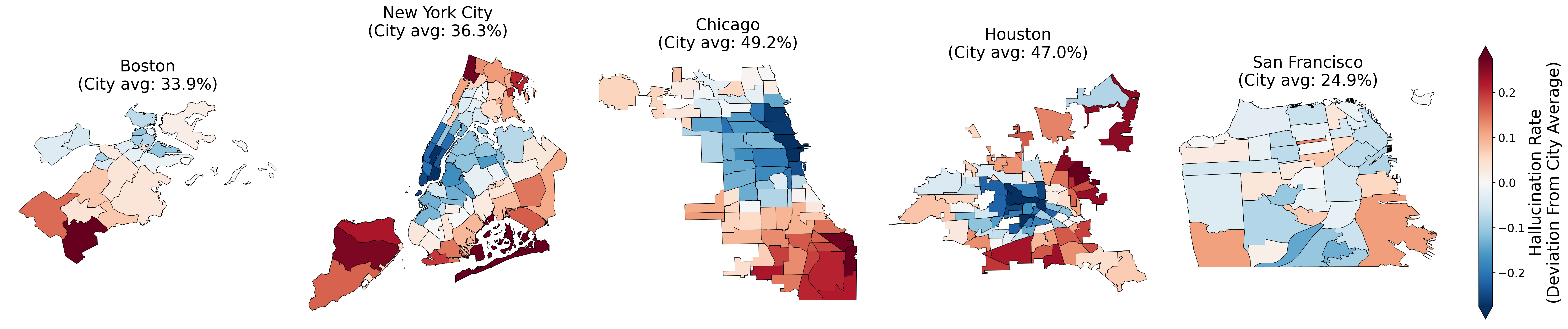}
    \caption{Neighborhood hallucination rate of open-ended venue recommendation by \texttt{Llama-3.3-70b-instruct}, mapped as the deviation from each city’s average.}
    \label{fig1d_llama-3_3-70b-instruct}
\end{figure}

\begin{figure}[ht]
    \centering
    \includegraphics[width=\linewidth]{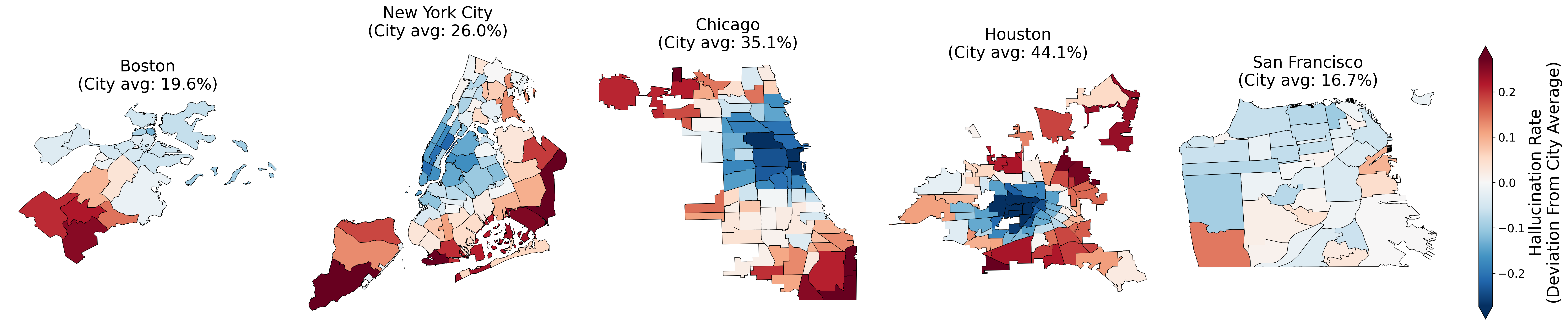}
    \caption{Neighborhood hallucination rate of open-ended venue recommendation by \texttt{Gemini-2.5-flash}, mapped as the deviation from each city’s average.}
    \label{fig1d_gemini-2_5-flash}
\end{figure}

\begin{figure}[ht]
    \centering
    \includegraphics[width=\linewidth]{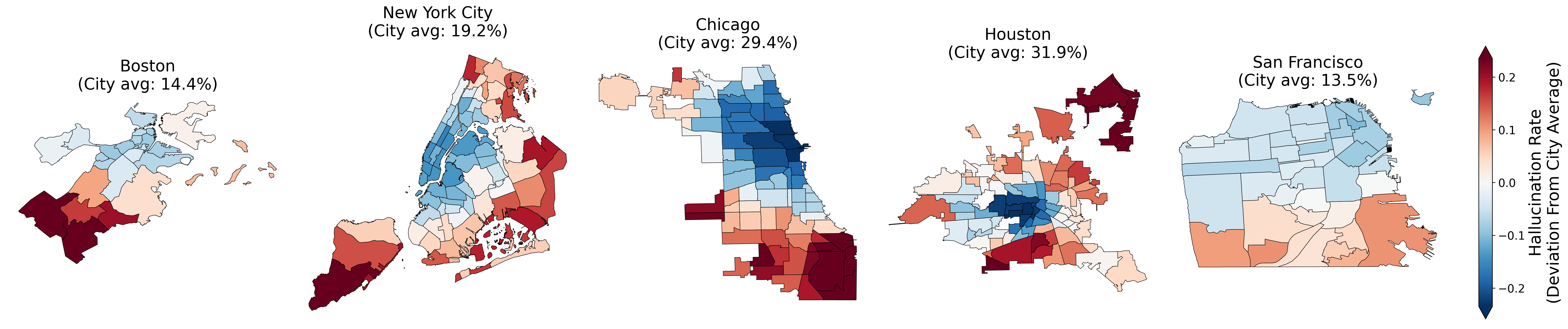}
    \caption{Neighborhood hallucination rate of open-ended venue recommendation by \texttt{GPT-6 Luna}, mapped as the deviation from each city’s average.}
    \label{fig1d_gpt-6-luna}
\end{figure}

\begin{figure}[ht]
    \centering
    \includegraphics[width=\linewidth]{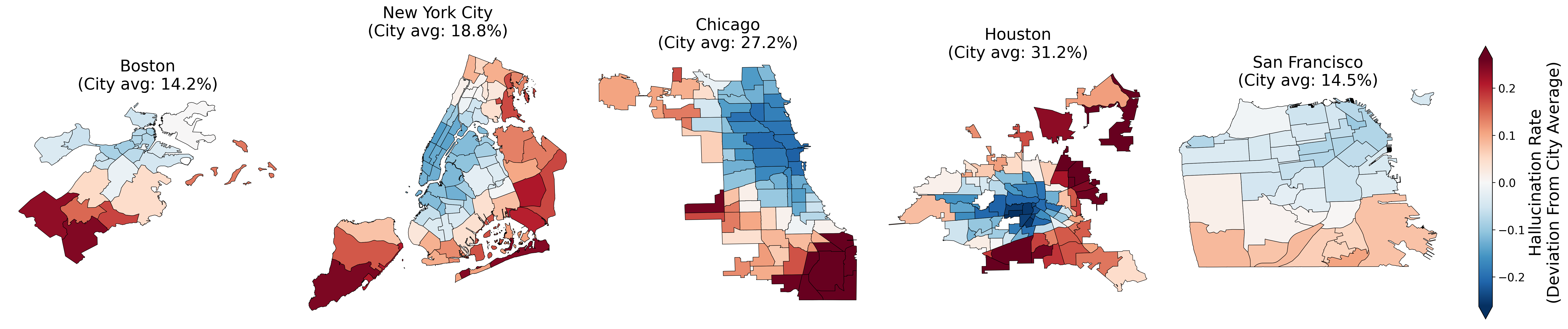}
    \caption{Neighborhood hallucination rate of open-ended venue recommendation by \texttt{Gemini-3.8-flash}, mapped as the deviation from each city’s average.}
    \label{fig1d_gemini-3_8-flash}
\end{figure}

\begin{figure}[ht]
    \centering
    \includegraphics[width=0.7\linewidth]{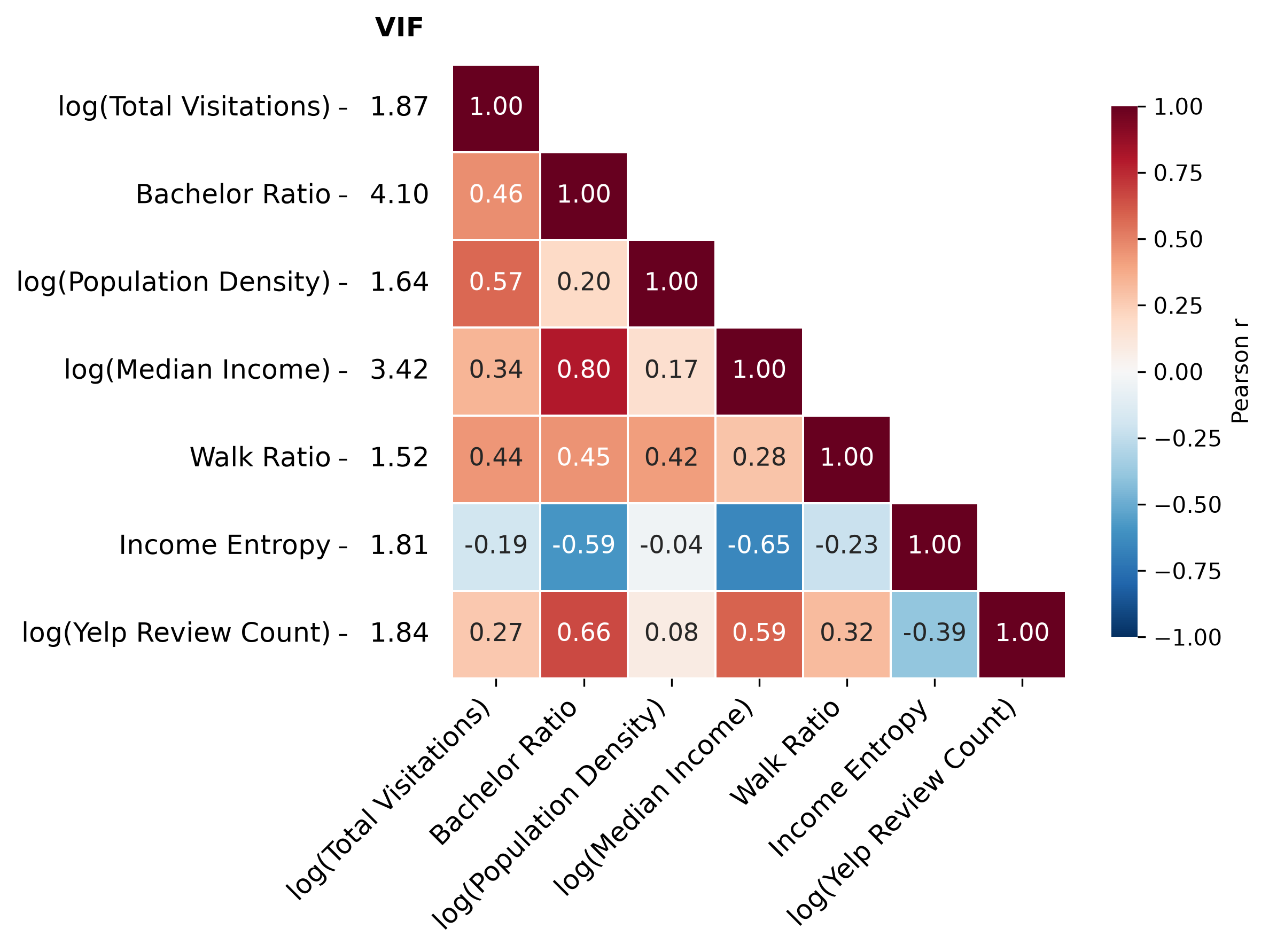}
    \caption{Neighborhood feature correlation matrix for regression covariates. Pearson correlations are shown for the neighborhood-level covariates used in the pooled regression models after applying log transformations to right-skewed variables. The left column reports the variance inflation factor (VIF) for each covariate, computed using the same set of regression features, to assess multicollinearity among predictors.}
    \label{fig1-2_nbh_feat_corr_matrix}
\end{figure}

\begin{figure}[ht]
    \centering
    \includegraphics[width=\linewidth]{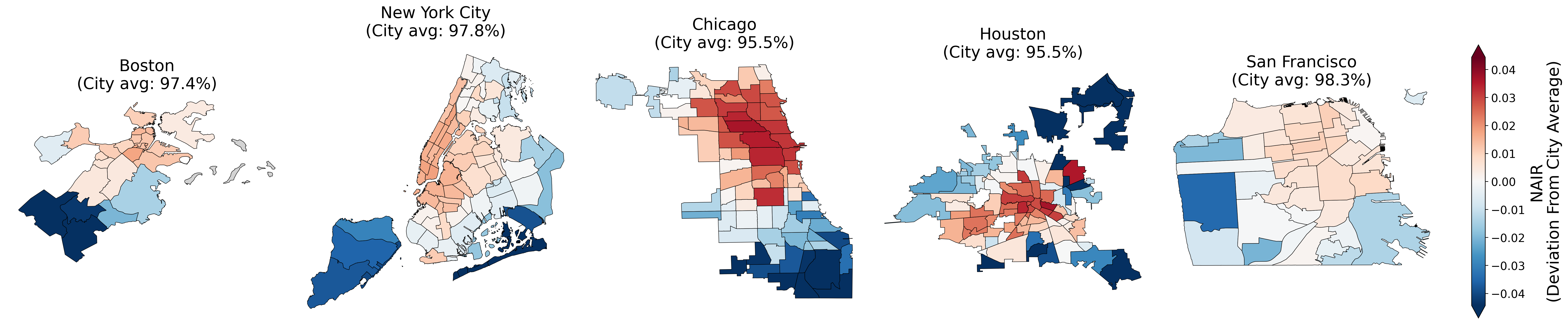}
    \caption{NAIR of open-ended venue recommendation by \texttt{Llama-3.3-70b-instruct}, mapped as the deviation from each city’s average.}
    \label{fig1e_llama-3_3-70b-instruct}
\end{figure}

\begin{figure}[ht]
    \centering
    \includegraphics[width=\linewidth]{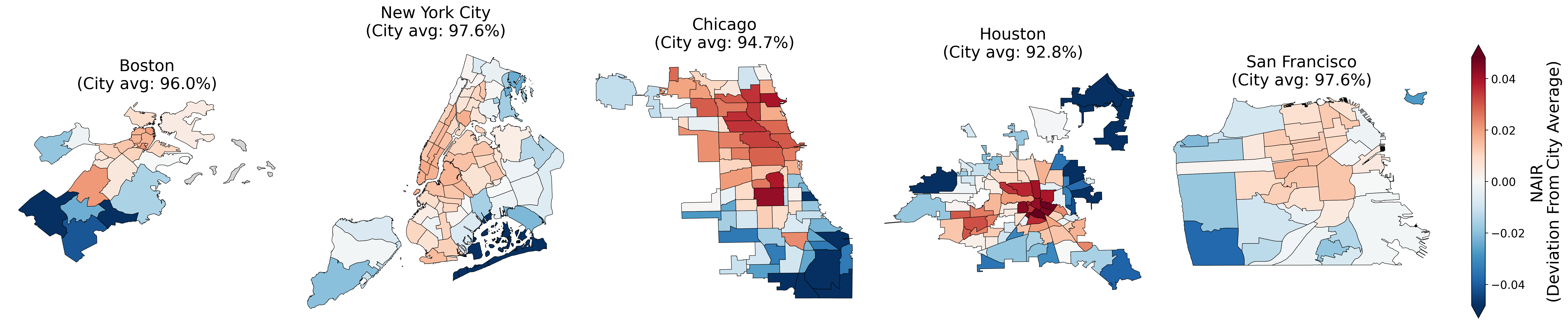}
    \caption{NAIR of open-ended venue recommendation by \texttt{Gemini-2.5-flash}, mapped as the deviation from each city’s average.}
    \label{fig1e_gemini-2_5-flash}
\end{figure}

\begin{figure}[ht]
    \centering
    \includegraphics[width=\linewidth]{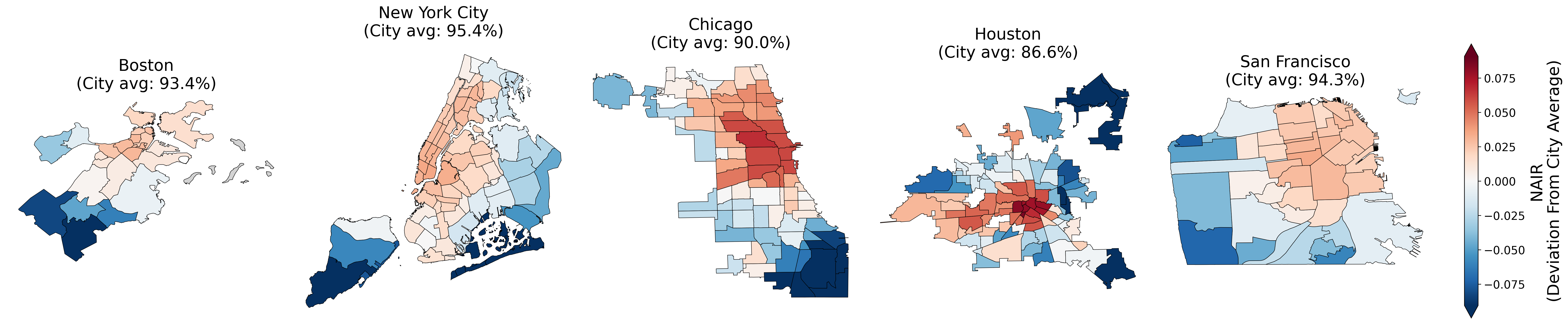}
    \caption{NAIR of open-ended venue recommendation by \texttt{GPT-6 Luna}, mapped as the deviation from each city’s average.}
    \label{fig1e_gpt-6-luna}
\end{figure}

\begin{figure}[ht]
    \centering
    \includegraphics[width=\linewidth]{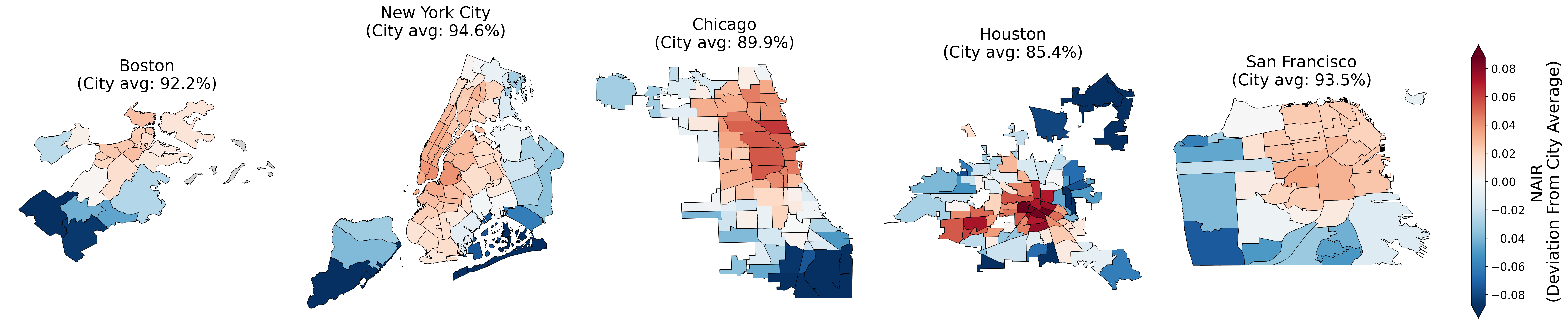}
    \caption{NAIR of open-ended venue recommendation by \texttt{Gemini-3.8-flash}, mapped as the deviation from each city’s average.}
    \label{fig1e_gemini-3_8-flash}
\end{figure}

\begin{figure}[ht]
    \centering
    \includegraphics[width=0.9\linewidth]{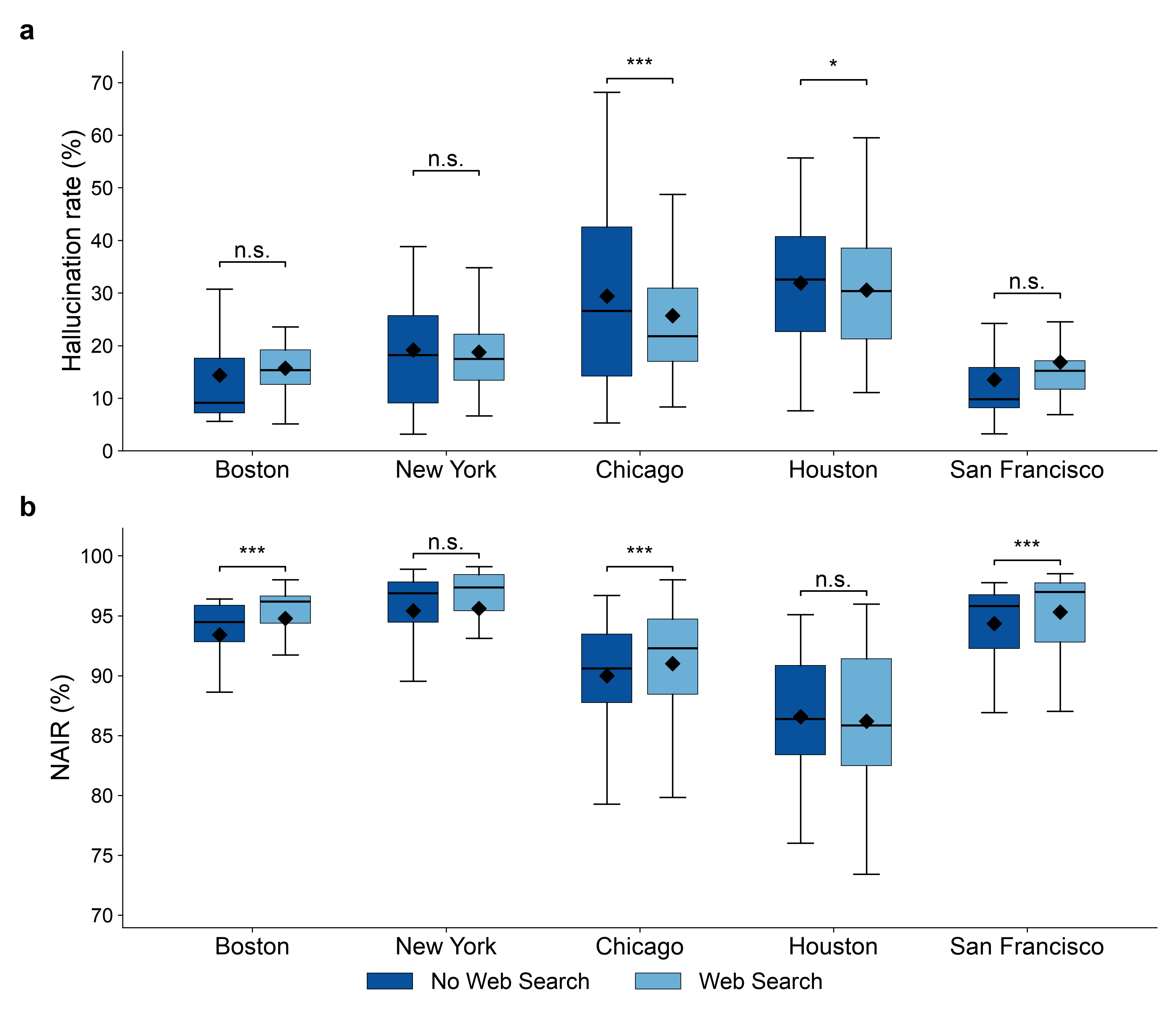}
    \caption{\rev{Comparison of GPT-6 Luna with and without web search in the open-ended setting, by city. \textbf{a}, Hallucination rate; \textbf{b}, NAIR. Box plots show the distribution across neighborhoods: median (center line), interquartile range (box), and 1.5$\times$ IQR (whiskers); diamonds mark the means. Brackets report one-sided paired $t$-tests across neighborhoods, testing whether the web-search variant has a lower hallucination rate (\textbf{a}) or a higher NAIR (\textbf{b}) ($^{***}p<0.001$, $^{**}p<0.01$, $^{*}p<0.05$; n.s., not significant).}}
    \label{fig:websearch_box}
\end{figure}

\begin{figure}[ht]
    \centering
    \includegraphics[width=\linewidth]{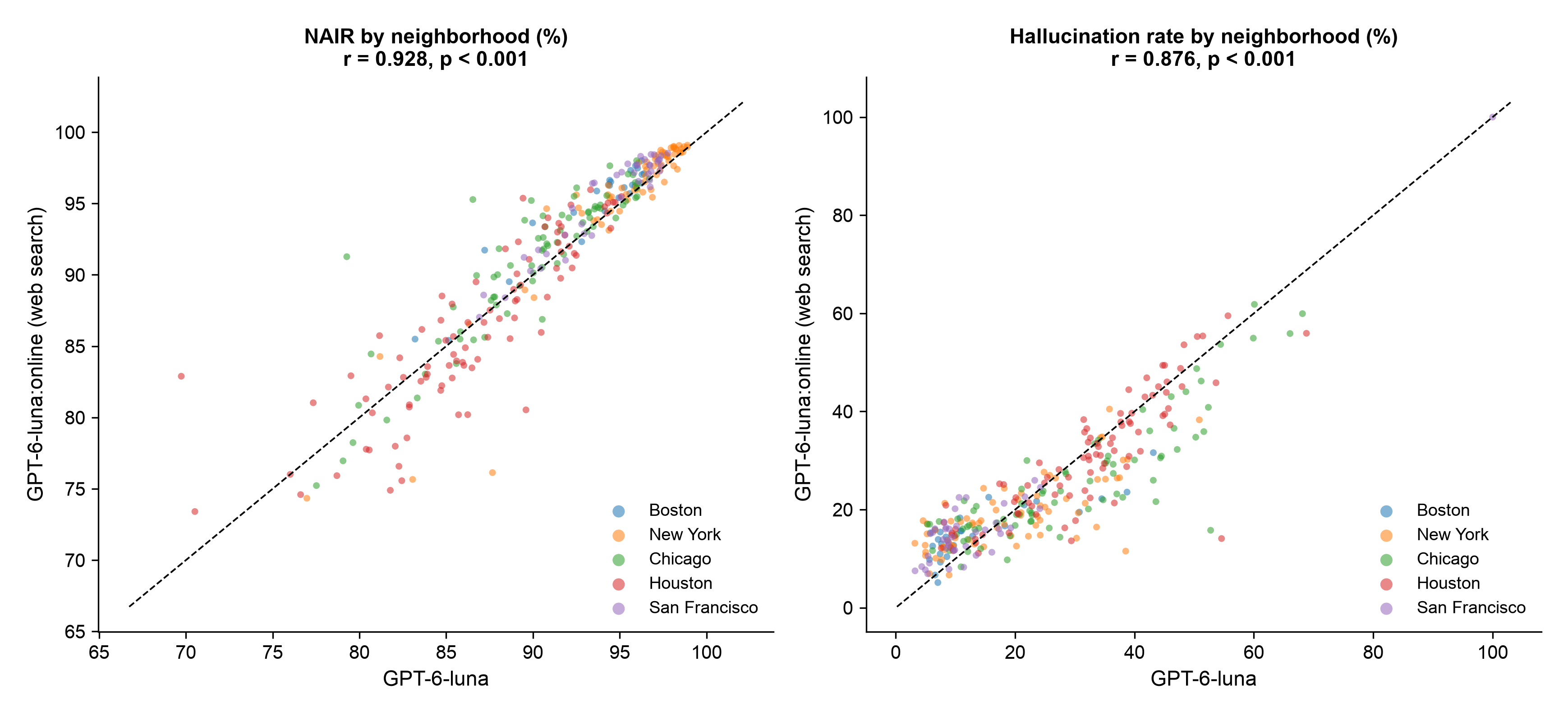}
    \caption{\rev{Neighborhood-level comparison of GPT-6 Luna with and without web search in the open-ended setting. Left, NAIR; right, hallucination rate. Each point represents a neighborhood, colored by city. The horizontal axis shows GPT-6 Luna without web search and the vertical axis shows its web-search-enabled variant (\texttt{gpt-6-luna:online}); the dashed line marks equality. Pearson correlation coefficients are reported above each panel.}}
    \label{fig:websearch}
\end{figure}

\begin{figure}[ht]
    \centering
    \includegraphics[width=\linewidth]{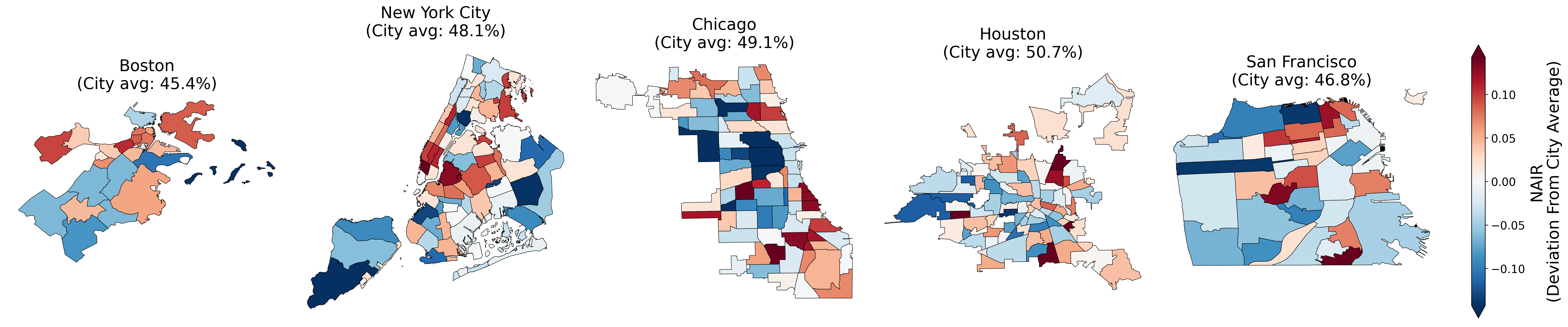}
    \caption{NAIR of candidate-constrained venue recommendation by \texttt{Llama-3.3-70b-instruct}, mapped as the deviation from each city’s average.}
    \label{fig2c_llama-3_3-70b-instruct}
\end{figure}

\begin{figure}[ht]
    \centering
    \includegraphics[width=\linewidth]{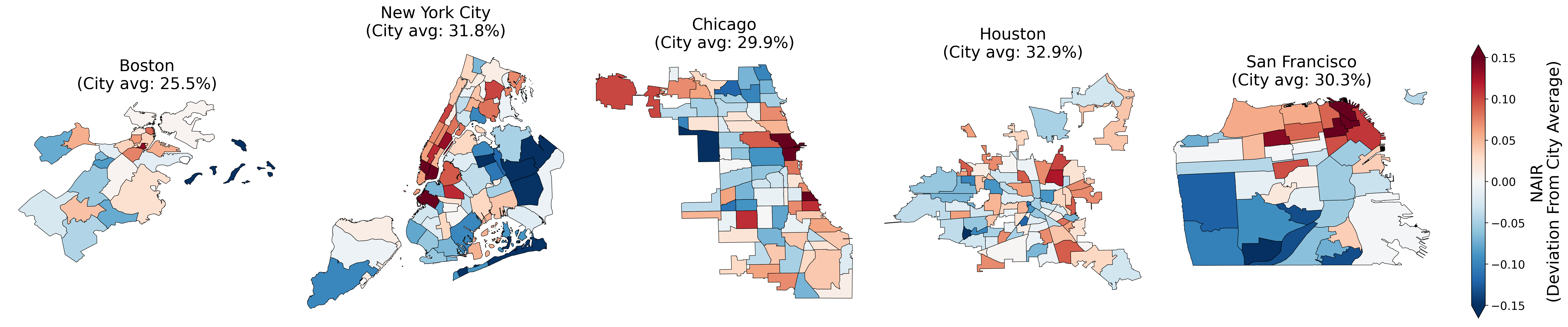}
    \caption{NAIR of candidate-constrained venue recommendation by \texttt{Gemini-2.5-flash}, mapped as the deviation from each city’s average.}
    \label{fig2c_gemini-2_5-flash}
\end{figure}

\begin{figure}[ht]
    \centering
    \includegraphics[width=\linewidth]{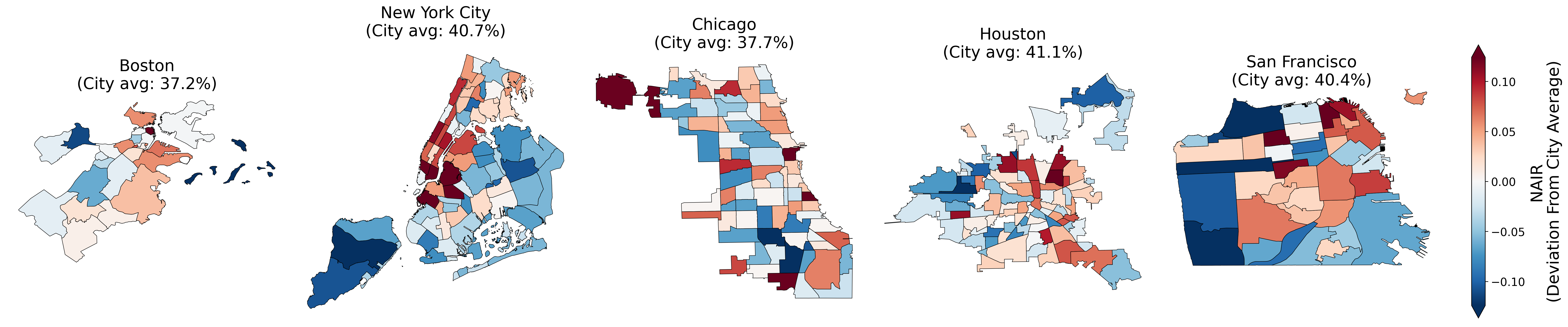}
    \caption{NAIR of candidate-constrained venue recommendation by \texttt{GPT-6 Luna}, mapped as the deviation from each city’s average.}
    \label{fig2c_gpt-6-luna}
\end{figure}

\begin{figure}[ht]
    \centering
    \includegraphics[width=\linewidth]{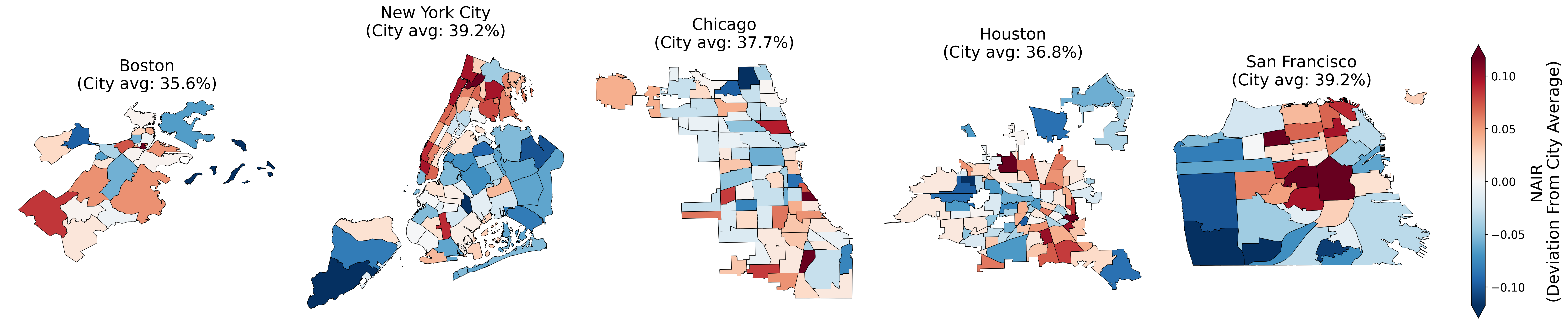}
    \caption{NAIR of candidate-constrained venue recommendation by \texttt{Gemini-3.8-flash}, mapped as the deviation from each city’s average.}
    \label{fig2c_gemini-3_8-flash}
\end{figure}

\begin{figure}[ht]
\centering
\includegraphics[width=\linewidth]{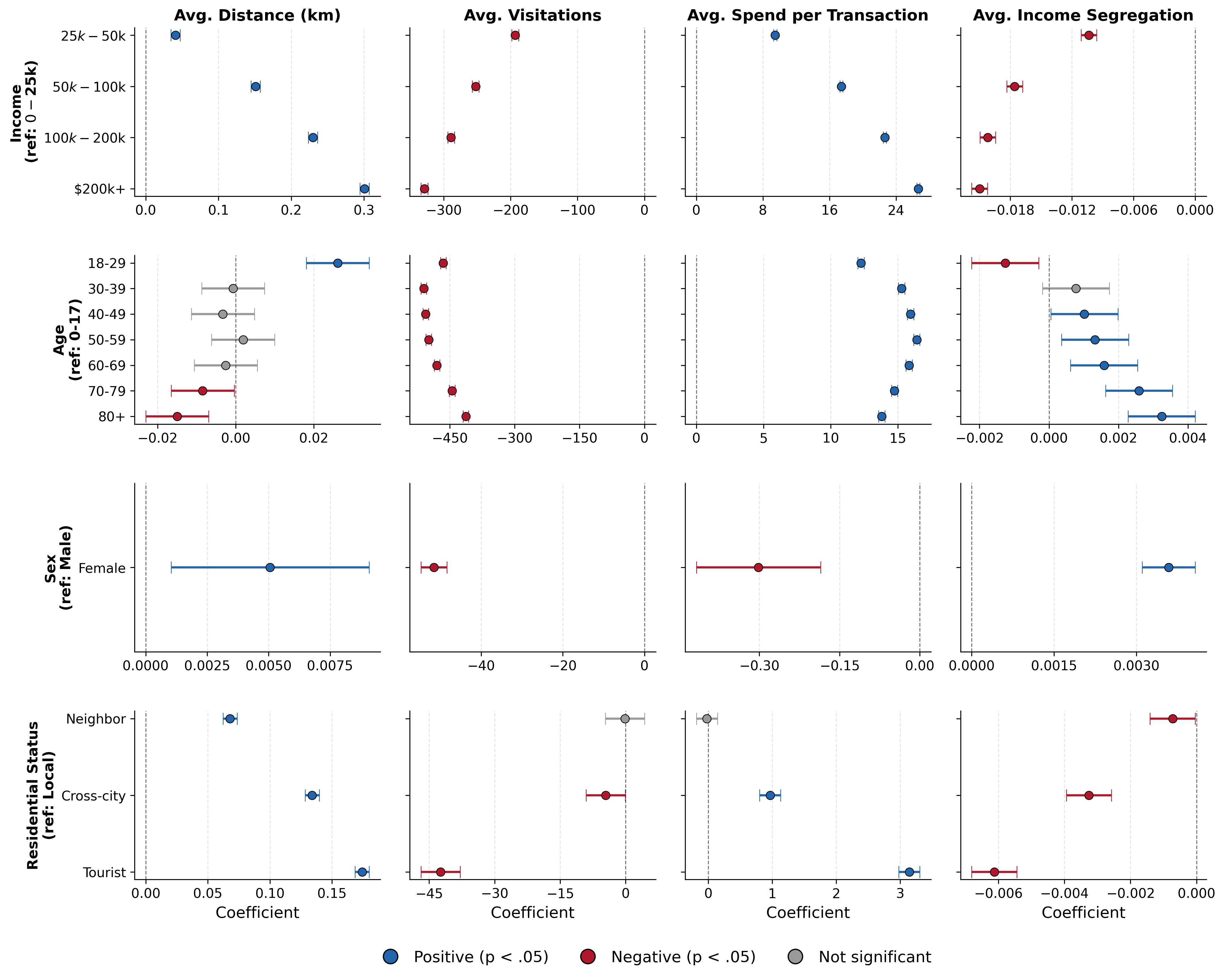}
\caption{\textbf{LLM-based venue recommendations \rev{differ systematically across demographic profiles}.} 
Mixed-effects regression coefficients for four outcome variables (columns) across four demographic dimensions (rows): \textbf{a}, income (reference: \$0–25k), \textbf{b}, age (reference: \rev{0–17}), \textbf{c}, sex (reference: male), and \textbf{d}, residential status (reference: local).
\rev{All regression models include LLM-model fixed effects and a city-level random intercept.}
Points show regression coefficients; horizontal whiskers denote \rev{95\% confidence intervals}.
Colors denote statistical significance (red: significantly negative ($p<0.05$); blue: significantly positive ($p<0.05$); grey: not significant ($p \geq 0.05$)).
}
\label{fig3_regression}
\end{figure}

\begin{figure}[ht]
    \centering
    \includegraphics[width=\linewidth]{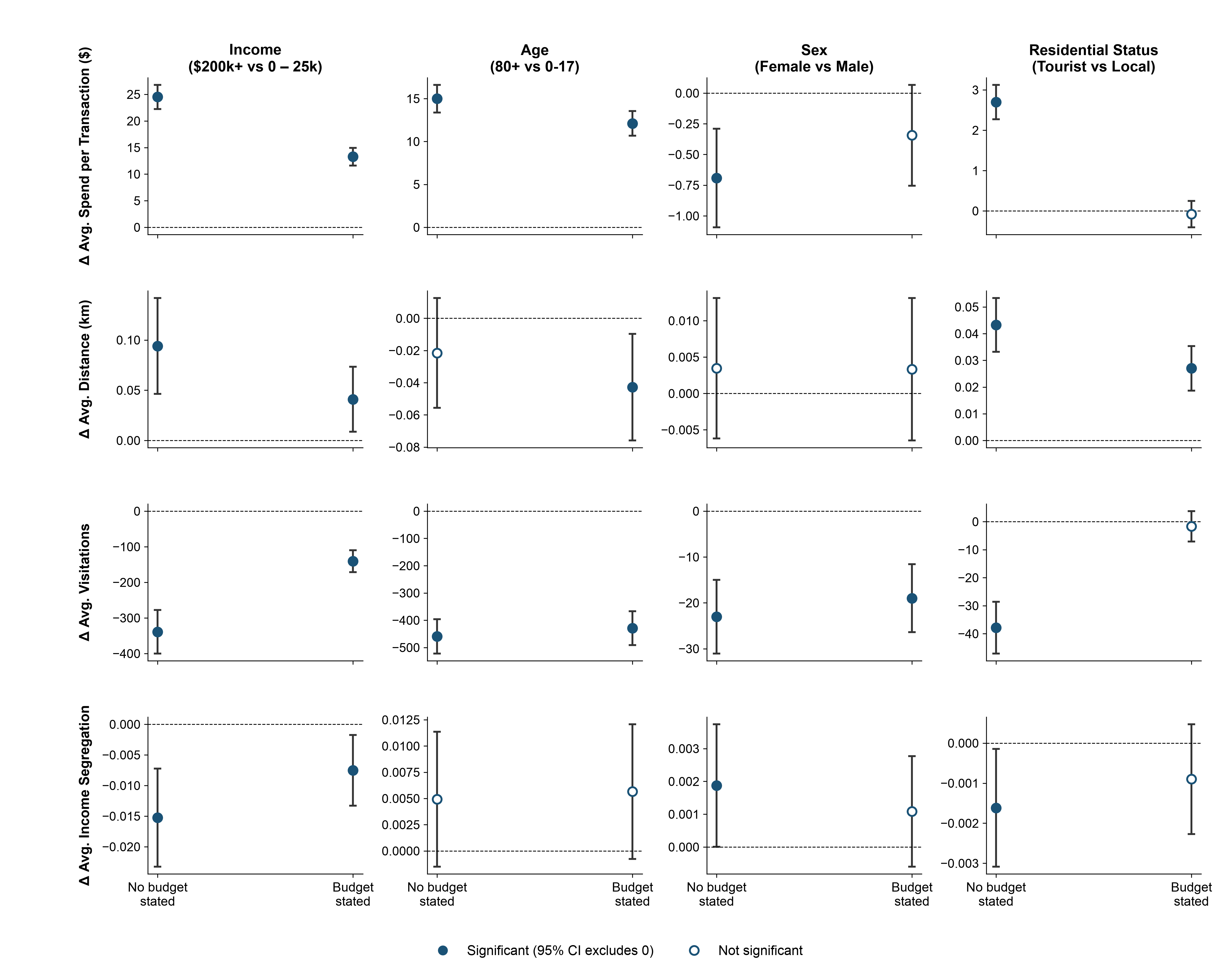}
    \caption{\rev{Demographic contrasts in recommended venue characteristics with and without a stated budget, using recommendations from \texttt{GPT-4o-mini}. Columns show four demographic contrasts (income: \$200k+ vs.\ \$0–25k; age: 80+ vs.\ 0–17; sex: female vs.\ male; residential status: tourist vs.\ local), and rows show four characteristics of the recommended venues. Within each panel, the left and right points show the estimated contrast when no budget is stated and when the budget is stated in the prompt, respectively; whiskers indicate 95\% CI. Filled markers denote contrasts whose 95\% CI excludes zero; open markers denote non-significant contrasts.}}
    \label{fig:budget}
\end{figure}

\begin{figure}[ht]
    \centering
    \includegraphics[width=0.65\linewidth]{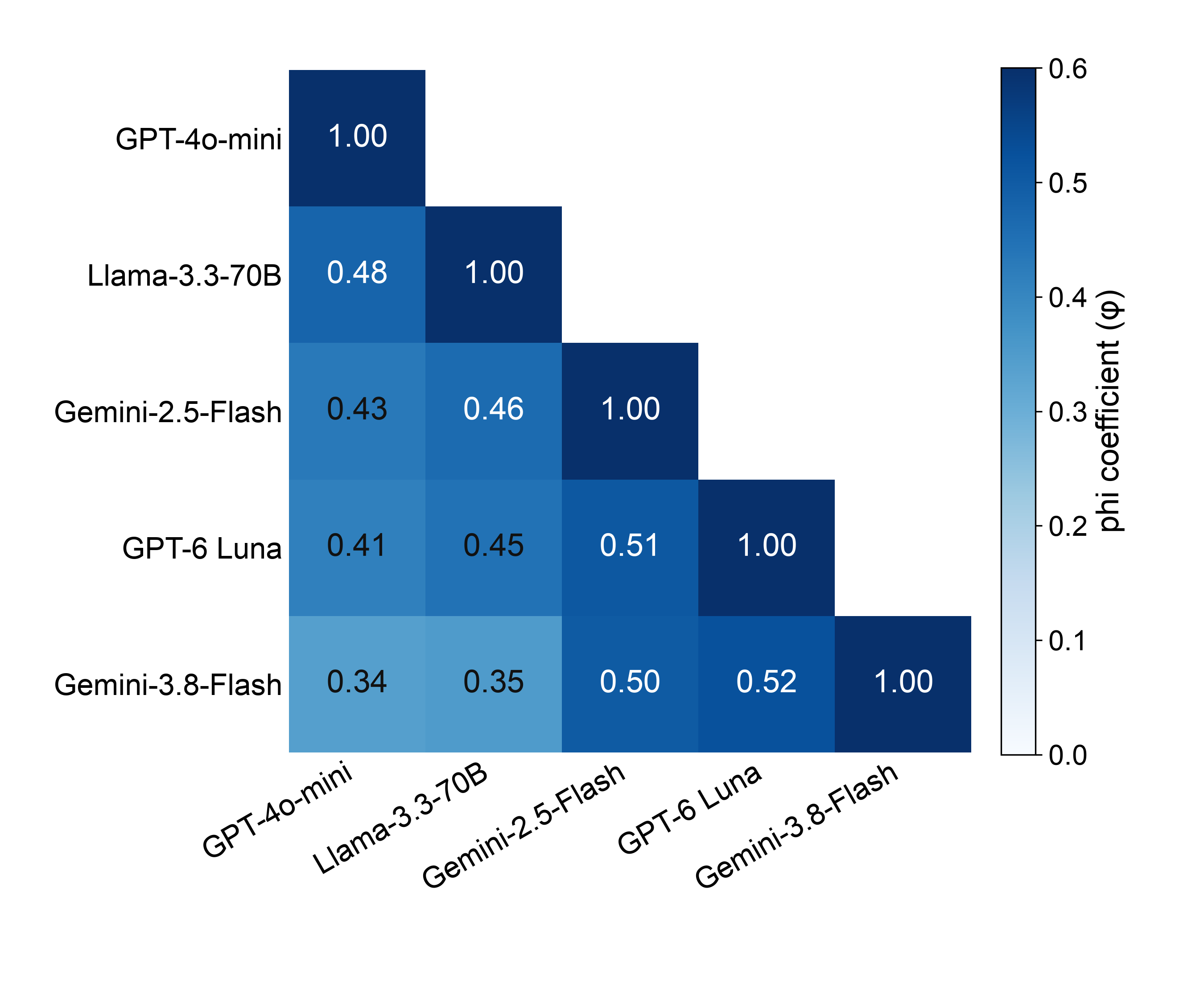}
    \caption{\rev{Pairwise association of venue silence between models in the candidate-constrained setting. Each cell reports the Phi coefficient between two models' binary indicators of whether a candidate venue receives no recommendation across all profiles. Darker shades indicate larger Phi coefficients, that is, a stronger tendency for the two models to overlook the same venues.}}
    \label{fig:phi}
\end{figure}

\begin{figure}[ht]
    \centering
    \includegraphics[width=\linewidth]{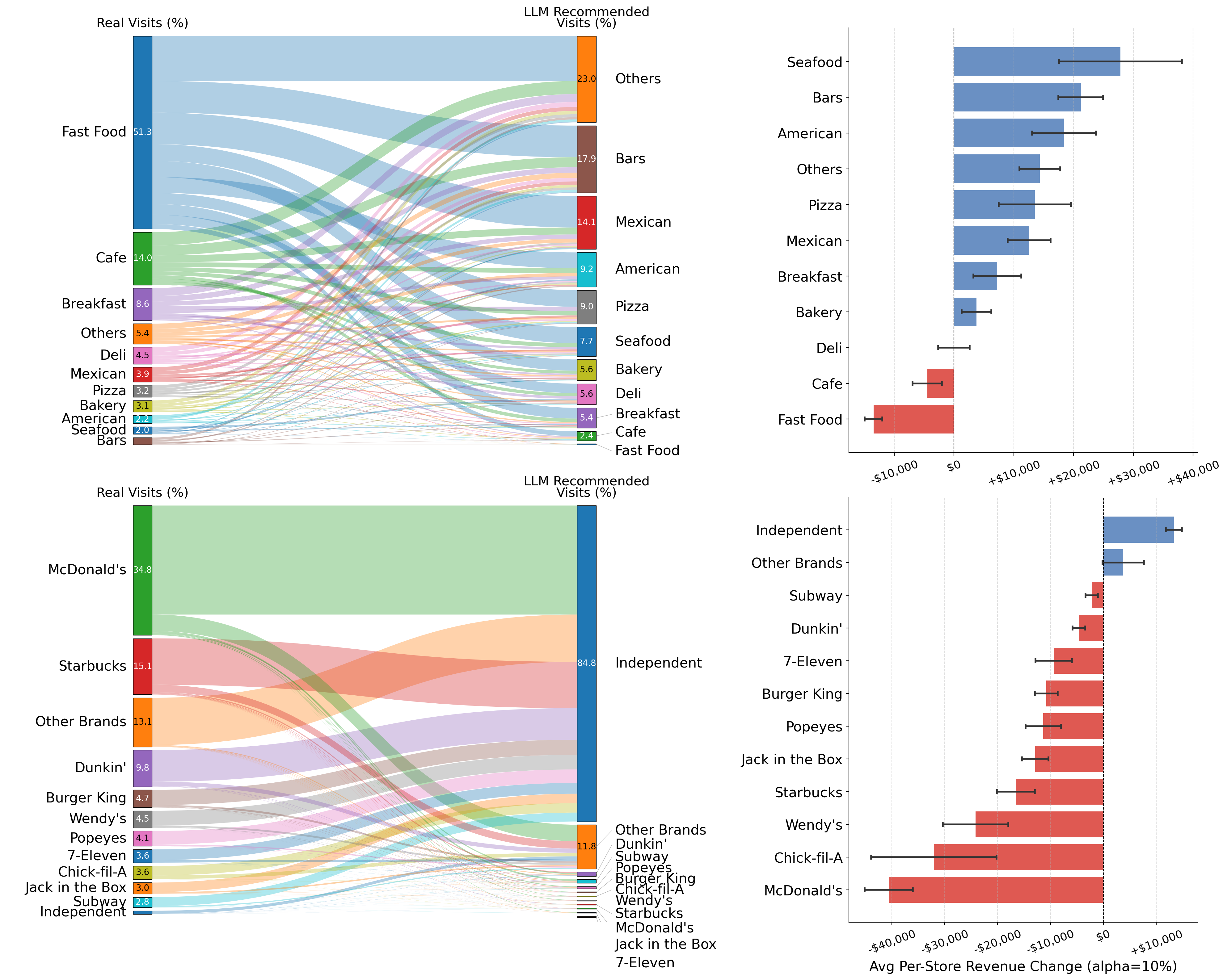}
    \caption{Redistribution of visitations and revenue across food categories and brands, when 10\% of visitations follow \texttt{GPT-4o-mini}.}
    \label{fig4_gpt-4o-mini}
\end{figure}

\begin{figure}[ht]
    \centering
    \includegraphics[width=\linewidth]{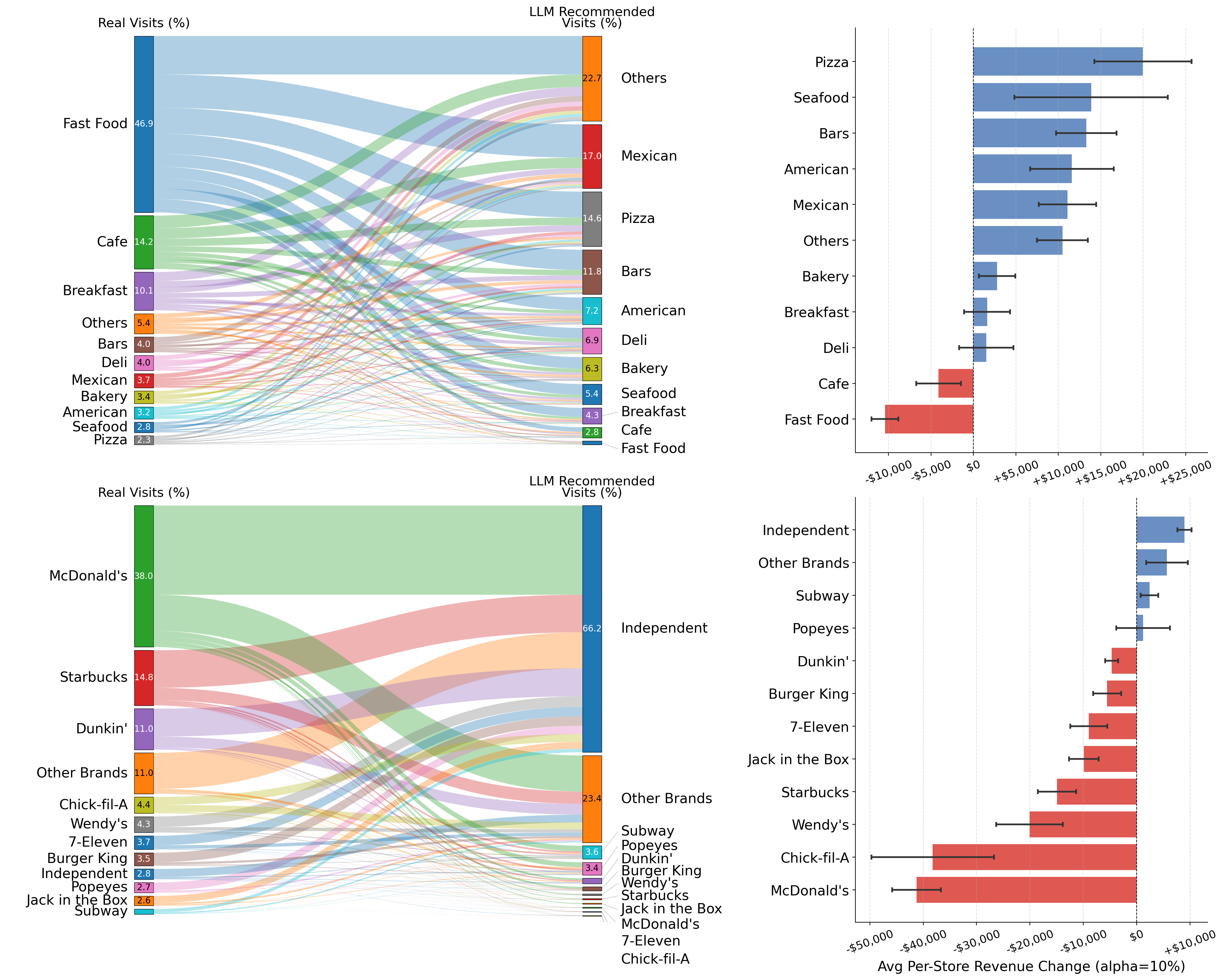}
    \caption{Redistribution of visitations and revenue across food categories and brands, when 10\% of visitations follow \texttt{Llama-3.3-70b-instruct}.}
    \label{fig4_llama-3_3-70b-instruct}
\end{figure}

\begin{figure}[ht]
    \centering
    \includegraphics[width=\linewidth]{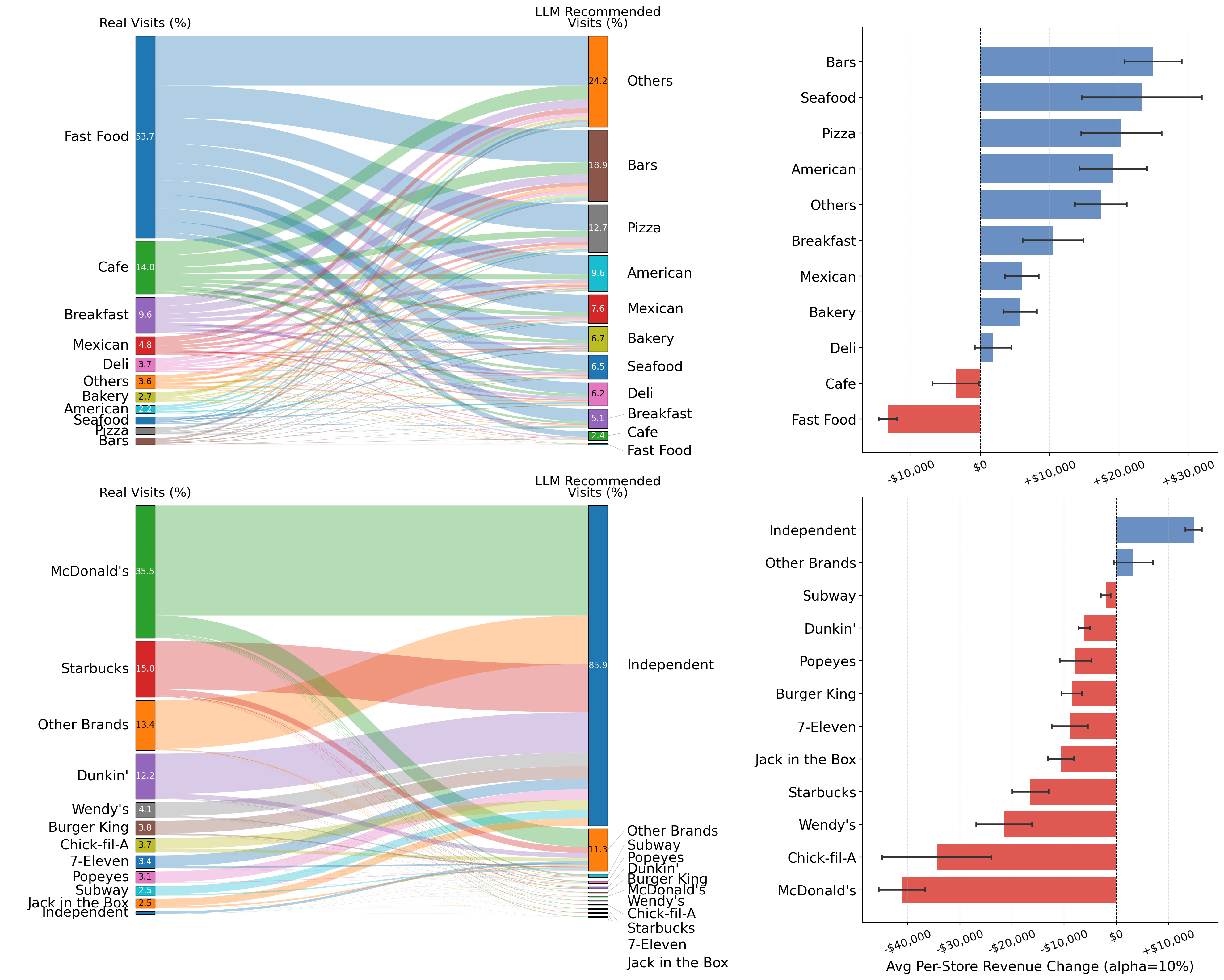}
    \caption{Redistribution of visitations and revenue across food categories and brands, when 10\% of visitations follow \texttt{Gemini-2.5-flash}.}
    \label{fig4_gemini-2_5-flash}
\end{figure}

\begin{figure}[ht]
    \centering
    \includegraphics[width=\linewidth]{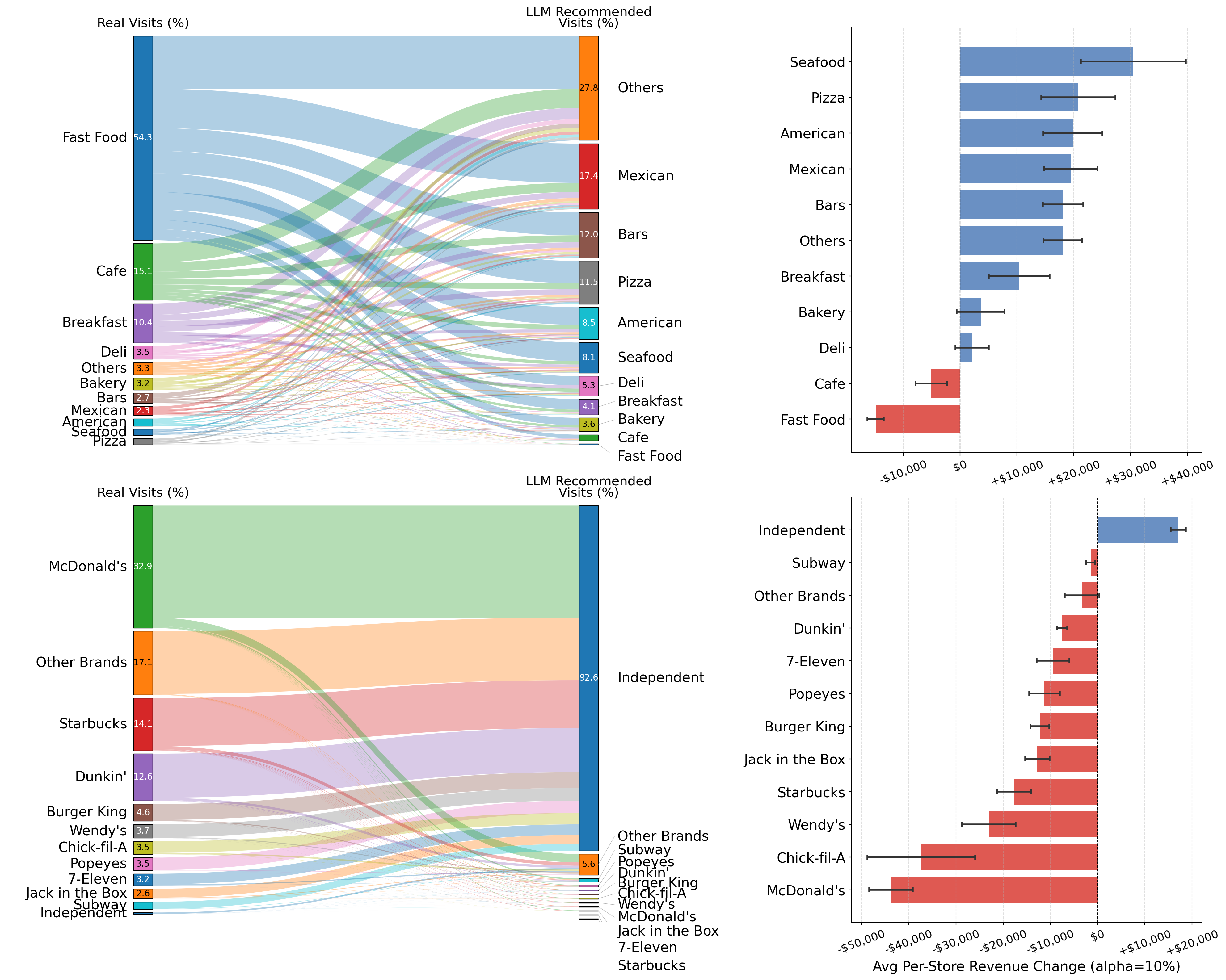}
    \caption{Redistribution of visitations and revenue across food categories and brands, when 10\% of visitations follow \texttt{GPT-6 Luna}.}
    \label{fig4_gpt-6-luna}
\end{figure}

\begin{figure}[ht]
    \centering
    \includegraphics[width=\linewidth]{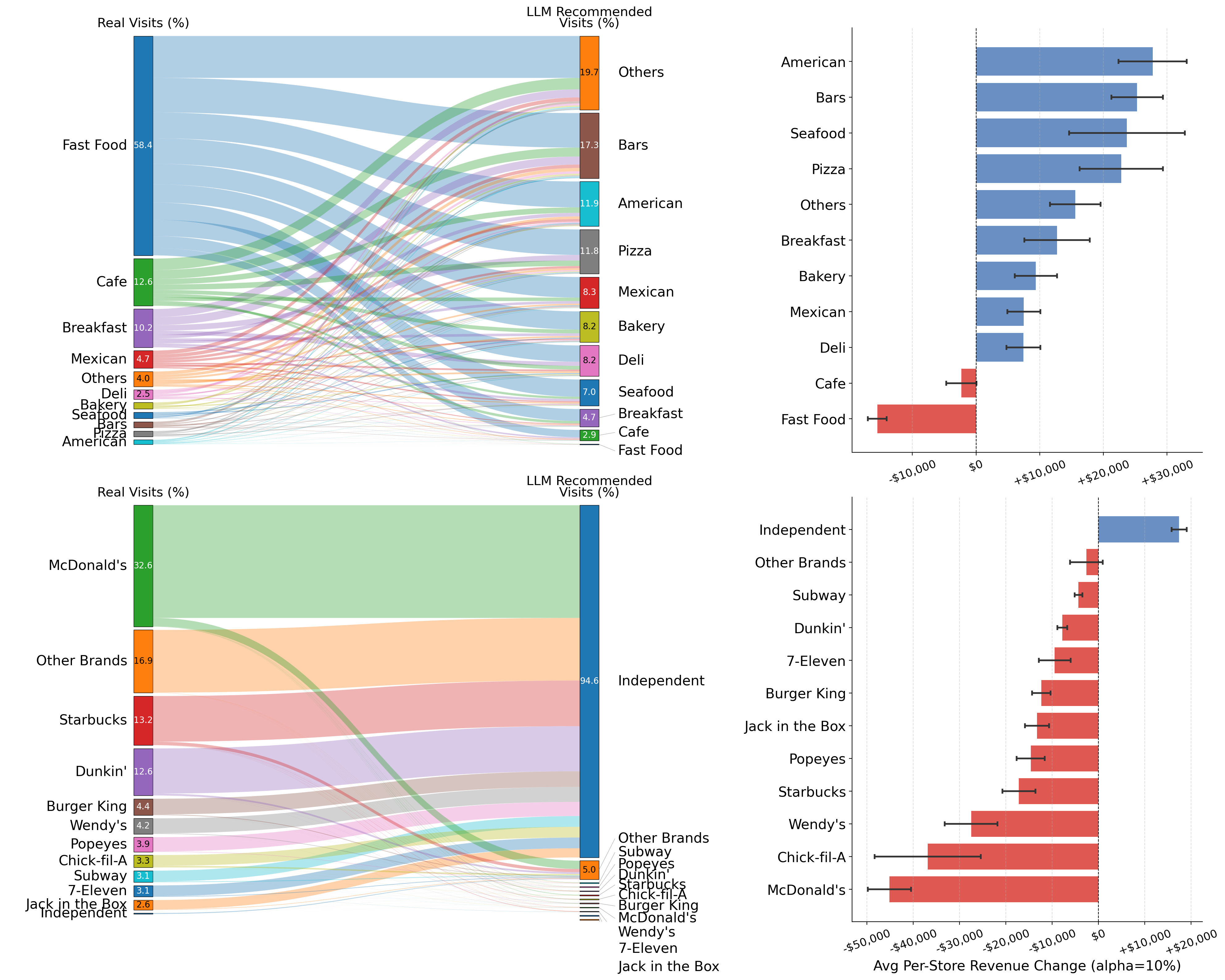}
    \caption{Redistribution of visitations and revenue across food categories and brands, when 10\% of visitations follow \texttt{Gemini-3.8-flash}.}
    \label{fig4_gemini-3_8-flash}
\end{figure}

\begin{figure}[ht]
    \centering
    \includegraphics[width=\linewidth]{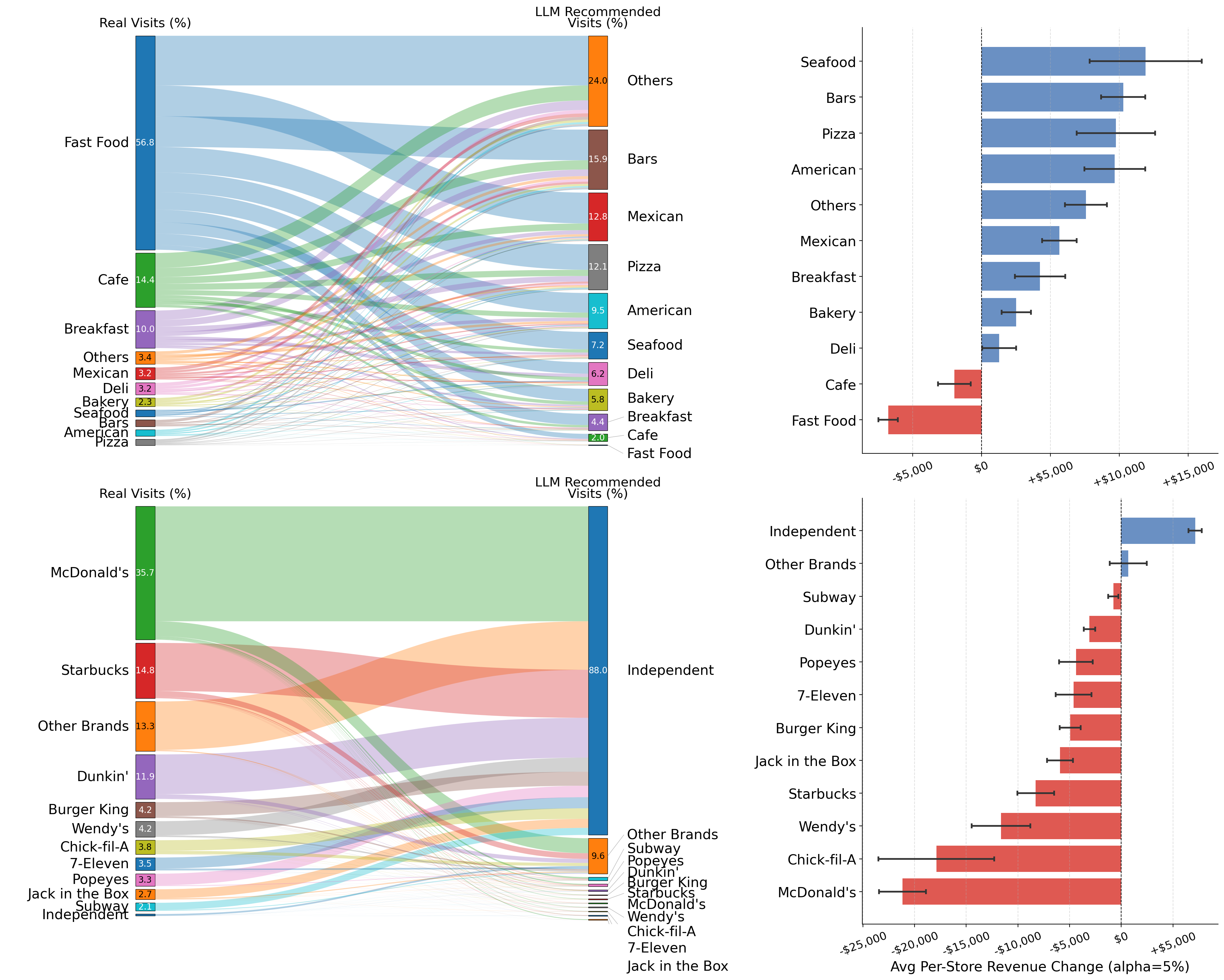}
    \caption{Redistribution of visitations and revenue across food categories and brands, when 5\% of visitations follow LLM recommendations.}
    \label{fig4_5percent}
\end{figure}

\begin{figure}[ht]
    \centering
    \includegraphics[width=\linewidth]{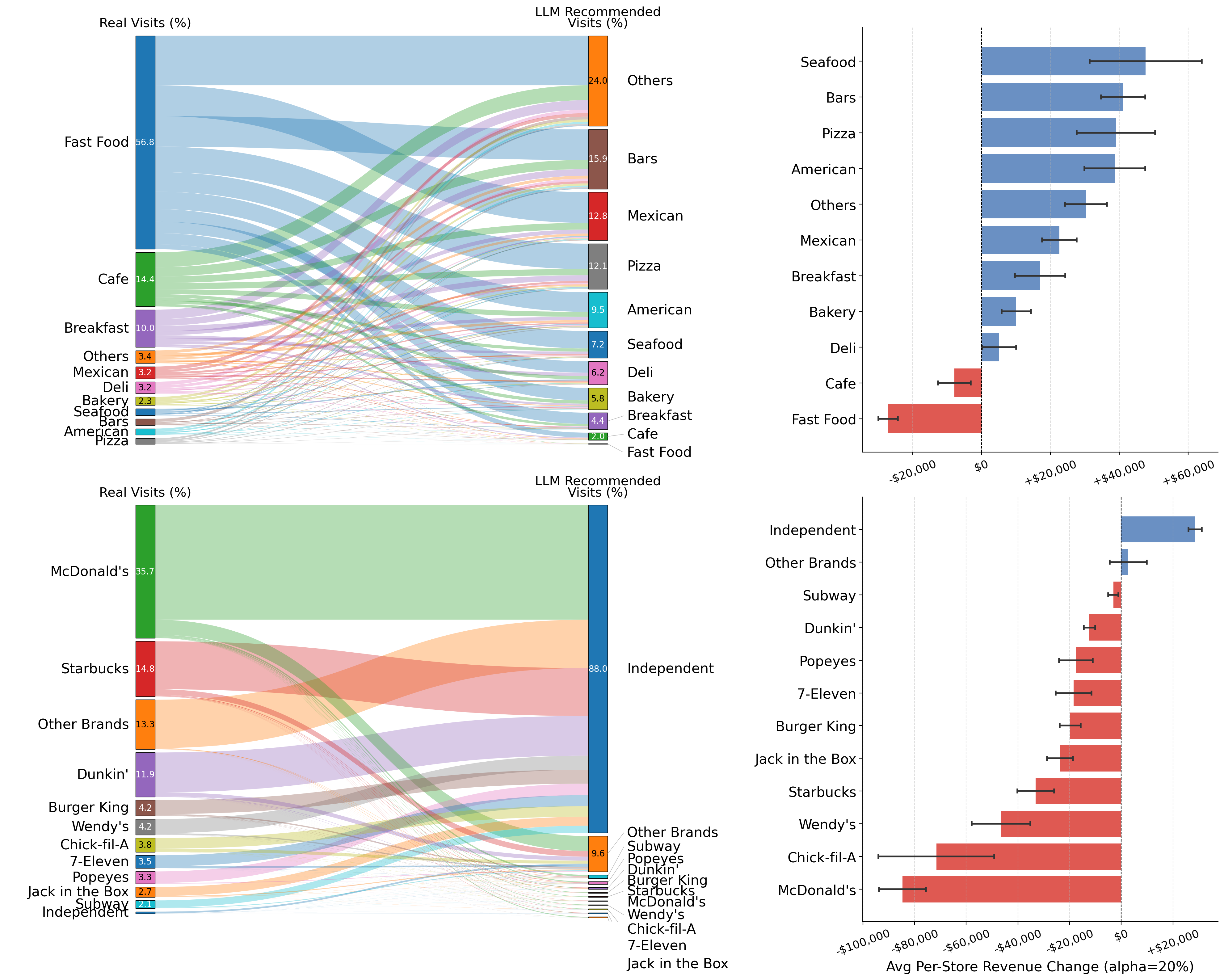}
    \caption{Redistribution of visitations and revenue across food categories and brands, when 20\% of visitations follow LLM recommendations.}
    \label{fig4_20percent}
\end{figure}

\begin{figure}[ht]
    \centering
    \includegraphics[width=\linewidth]{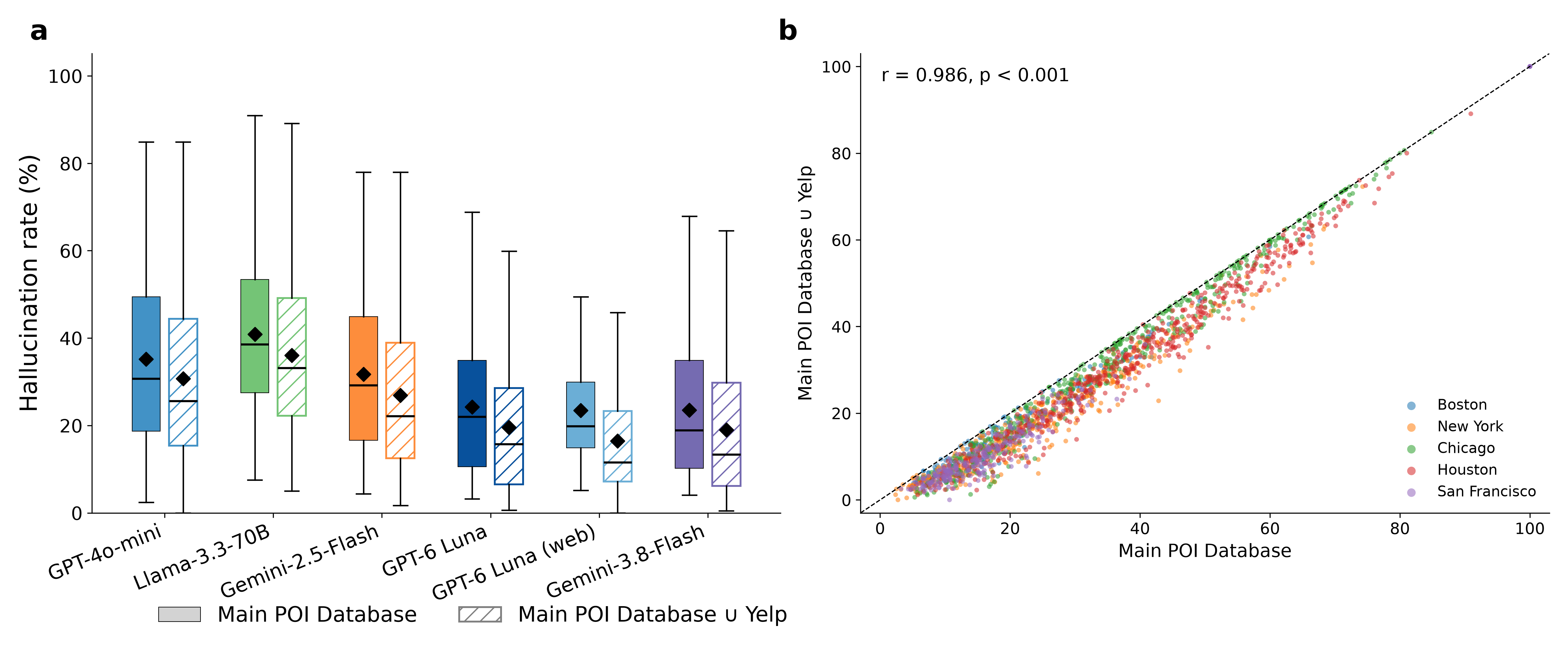}
    \caption{\rev{\textbf{Hallucination rates under the main POI database and under its union with Yelp POI data.} \textbf{a}, Distribution of neighborhood hallucination rates for each model when LLM outputs are matched against the main POI database (solid boxes) and against the union of the main POI database and Yelp POI data (hatched boxes). Box plots show median (center line), interquartile range (box), and 1.5× IQR (whiskers); diamonds indicate means. \textbf{b}, Neighborhood hallucination rates under the two reference sets. Each point represents one neighborhood–model pair, and colors indicate cities. The dashed line marks equality, and $r$ denotes the Pearson correlation coefficient.}}
    \label{fig:hallu_yelp}
\end{figure}

\clearpage

\begin{table}[ht]\centering
\caption{Pooled Regression of Hallucination Rate on Neighborhood Features}
\label{tab:hallucination-rate-pooled-regression}
\begin{tabular}{lcccc}
\hline
Variable & Coefficient & Std. Error & p-value & 95\% CI \\
\hline
Bachelor Ratio & -0.053$^{***}$ & (0.005) & 0.000 & [-0.063, -0.042] \\
log(Total Visitations) & -0.098$^{***}$ & (0.004) & 0.000 & [-0.106, -0.090] \\
log(Population Density) & -0.013$^{**}$ & (0.004) & 0.003 & [-0.021, -0.004] \\
log(Median Income) & 0.044$^{***}$ & (0.005) & 0.000 & [0.034, 0.054] \\
log(Walk Ratio) & 0.018$^{***}$ & (0.004) & 0.000 & [0.011, 0.026] \\
Income Entropy & -0.001 & (0.004) & 0.743 & [-0.009, 0.006] \\
log(Yelp Review Count) & -0.029$^{***}$ & (0.005) & 0.000 & [-0.038, -0.020] \\
\hline
Controls & None & & & \\
City FE & Yes & & & \\
Model FE & Yes & & & \\
Observations & 1,430 & & & \\
$R^2$ & 0.744 & & & \\
Adjusted $R^2$ & 0.741 & & & \\
\hline
\multicolumn{5}{l}{\footnotesize Notes: Coefficients are standardized. Standard errors are in parentheses.} \\
\multicolumn{5}{l}{\footnotesize $^{***}p<0.001$, $^{**}p<0.01$, $^{*}p<0.05$.} \\
\end{tabular}
\end{table}

\begin{table}[ht]\centering
\caption{Pooled Regression of NAIR on Neighborhood Features}
\label{tab:silent-ratio-pooled-regression}
\begin{tabular}{lcccc}
\hline
Variable & Coefficient & Std. Error & p-value & 95\% CI \\
\hline
Bachelor Ratio & -0.001 & (0.002) & 0.371 & [-0.004, 0.002] \\
log(Total Visitations) & -0.009$^{***}$ & (0.002) & 0.000 & [-0.013, -0.005] \\
log(Population Density) & -0.005$^{***}$ & (0.001) & 0.000 & [-0.007, -0.003] \\
log(Median Income) & -0.001 & (0.001) & 0.646 & [-0.003, 0.002] \\
log(Walk Ratio) & -0.001 & (0.001) & 0.146 & [-0.003, 0.001] \\
Income Entropy & 0.003$^{**}$ & (0.001) & 0.006 & [0.001, 0.005] \\
log(Yelp Review Count) & -0.004$^{***}$ & (0.001) & 0.000 & [-0.007, -0.002] \\
log(POI Count) & 0.052$^{***}$ & (0.003) & 0.000 & [0.046, 0.057] \\
\hline
Controls & log(POI Count) & & & \\
City FE & Yes & & & \\
Model FE & Yes & & & \\
Observations & 1,430 & & & \\
$R^2$ & 0.776 & & & \\
Adjusted $R^2$ & 0.774 & & & \\
\hline
\multicolumn{5}{l}{\footnotesize Notes: Coefficients are standardized. Standard errors are in parentheses.} \\
\multicolumn{5}{l}{\footnotesize $^{***}p<0.001$, $^{**}p<0.01$, $^{*}p<0.05$.} \\
\end{tabular}
\end{table}

\begin{table}[ht]\centering
\caption{Pooled Regression of Candidate-Constrained NAIR on Neighborhood Features}
\label{tab:silent-rate-pooled-regression-with-candidates}
\begin{tabular}{lcccc}
\hline
Variable & Coefficient & Std. Error & p-value & 95\% CI \\
\hline
Bachelor Ratio & 0.009$^{*}$ & (0.004) & 0.030 & [0.001, 0.016] \\
log(Total Visitations) & -0.007 & (0.006) & 0.184 & [-0.018, 0.004] \\
log(Population Density) & 0.019$^{***}$ & (0.003) & 0.000 & [0.013, 0.025] \\
log(Median Income) & 0.002 & (0.004) & 0.654 & [-0.006, 0.009] \\
log(Walk Ratio) & 0.008$^{**}$ & (0.003) & 0.002 & [0.003, 0.013] \\
Income Entropy & 0.000 & (0.003) & 0.901 & [-0.005, 0.005] \\
log(Yelp Review Count) & -0.016$^{***}$ & (0.003) & 0.000 & [-0.022, -0.010] \\
\hline
Controls & log(POI Count) & & & \\
City FE & Yes & & & \\
Model FE & Yes & & & \\
Observations & 1,430 & & & \\
$R^2$ & 0.576 & & & \\
Adjusted $R^2$ & 0.571 & & & \\
\hline
\multicolumn{5}{l}{\footnotesize Notes: Coefficients are standardized. Standard errors are in parentheses.} \\
\multicolumn{5}{l}{\footnotesize $^{***}p<0.001$, $^{**}p<0.01$, $^{*}p<0.05$.} \\
\end{tabular}
\end{table}

\begin{table}[ht]\centering
\caption{Pooled Logistic Regression Predicting Candidate POI Recommendation}
\label{tab:logit-recommended-pooled-r1}
\begin{tabular}{lccccc}
\hline
Variable & Coefficient & Std. Error & p-value & 95\% CI & Odds Ratio \\
\hline
Visit volume (log) & 0.114$^{***}$ & (0.010) & 0.000 & [0.094, 0.134] & 1.121 \\
Visitor income segregation & -0.030$^{**}$ & (0.009) & 0.001 & [-0.048, -0.011] & 0.971 \\
Median spend per transaction & 0.122$^{***}$ & (0.013) & 0.000 & [0.096, 0.148] & 1.129 \\
Yelp review count (log) & 0.330$^{***}$ & (0.010) & 0.000 & [0.310, 0.350] & 1.391 \\
Yelp rating & 0.023$^{*}$ & (0.010) & 0.019 & [0.004, 0.041] & 1.023 \\
\hline
City FE & Yes & & & & \\
Model FE & Yes & & & & \\
Observations & 60,660 & & & & \\
Pseudo $R^2$ & 0.048 & & & & \\
\hline
\multicolumn{6}{l}{\footnotesize Notes: Outcome equals one if a candidate POI is recommended.} \\
\multicolumn{6}{l}{\footnotesize Continuous predictors are standardized. Standard errors are in parentheses.} \\
\multicolumn{6}{l}{\footnotesize $^{***}p<0.001$, $^{**}p<0.01$, $^{*}p<0.05$.} \\
\end{tabular}
\end{table}

\small
\bibliographystyle{unsrt}
\bibliography{sample}

\let\bibliographystyle\oldbibliographystyle
\let\bibliography\oldbibliography

\end{document}